\documentclass[11pt,letterpaper]{article}
\usepackage{jheppub}

\usepackage{graphicx}
\usepackage{bbm}
\usepackage{amsmath}
\usepackage{amssymb}
\usepackage{mathtools}
\usepackage{amsthm}
\usepackage{mathrsfs}
\usepackage{marvosym}
\usepackage{dsfont}
\usepackage{tikz, tikz-cd, tikzsymbols}
\usepackage[labelformat=simple]{subcaption}
\usepackage{xcolor}
\usepackage{braket}
\usepackage{cleveref}
\usepackage{comment}
\usepackage{float}
\usepackage{fancybox}
\usepackage{slashed}
\usepackage[skins,theorems]{tcolorbox}
\tcbset{highlight math style={enhanced,
		colframe=black,colback=white,arc=0pt,boxrule=1pt}}
\usepackage{overpic}

\allowdisplaybreaks

\definecolor{dark-gray}{gray}{0.20}
\definecolor{gray}{gray}{0.30}
\definecolor{light-gray}{gray}{0.80}
\definecolor{dark-red}{rgb}{0.7,0,0}
\definecolor{dark-green}{rgb}{0.1,0.4,0}
\definecolor{dark-blue}{rgb}{0.3,0.3,0.7}
\definecolor{light-blue}{rgb}{0.8,0.8,1}
\definecolor{swamp}{RGB}{240, 199, 197}

\newcommand{\be}{\begin{equation}}
	\newcommand{\ee}{\end{equation}}

\def\be{\begin{equation}}
	\def\ee{\end{equation}}
\def\bea{\begin{eqnarray}}
	\def\eea{\end{eqnarray}}

\newcommand{\beq}{\begin{equation}}  \newcommand{\eeq}{\end{equation}}
\newcommand{\bal}{\begin{aligned}}   \newcommand{\eal}{\end{aligned}}
\def\beqa{\begin{eqnarray}}
	\def\eeqa{\end{eqnarray}}

\newcommand{\ii}{\mathrm i}
\newcommand{\cof}{\operatorname{cof}}

\newcommand{\tr}{\operatorname{tr}}
\newcommand{\vol}{\operatorname{vol}}

\newcommand{\N}{\mathcal N}
\newcommand{\Bh}{\mathcal B_h}
\newenvironment{eqn}{\begin{equation}\begin{aligned}}{\end{aligned}\end{equation}}
\newcommand{\G}{\boldsymbol\Gamma}
\newcommand{\Ham}{\mathcal H}
\newcommand{\Ga}{\mathcal G}

\numberwithin{equation}{section}

\newcommand{\Tr}{\mathrm{Tr}}

\numberwithin{equation}{section}

\def\simleq{\; \raise0.3ex\hbox{$<$\kern-0.75em
		\raise-1.1ex\hbox{$\sim$}}\; }
\def\simgeq{\; \raise0.3ex\hbox{$>$\kern-0.75em
		\raise-1.1ex\hbox{$\sim$}}\; }

\numberwithin{equation}{section}

\hypersetup{
	colorlinks=true,
	linkcolor=dark-blue,
	citecolor=dark-red,
	urlcolor=dark-green,
	linktoc=page
}

\theoremstyle{remark}

\newtheoremstyle{named}{}{}{\itshape}{}{\bfseries}{.}{.5em}{#3}
\theoremstyle{named}

\title{\centering\bf{A matrix model of an M5--brane \\ on a plane--wave background}}

\author{Carmine Montella}
\affiliation{
\textit{II. Institut f\"ur Theoretische Physik, Universit\"at Hamburg, Notkestrasse 9,\\ 22607 Hamburg, Germany} \\ [3mm]
\textit{Zentrum f\"ur Mathematische Physik, Universit\"at Hamburg, \\ 
Bundesstrasse 55, 20146 Hamburg, Germany}
}

\emailAdd{carmine.montella@uni-hamburg.de}

\abstract{
We construct the maximally supersymmetric deformation of the recently proposed $\mathcal A_6$ matrix model on a plane--wave background. We derive the same theory independently, by reducing the light--cone M5--brane on a five--sphere in this background to its lowest harmonics, including the dynamics of its chiral two--form and of its fermions. We then classify all its classical vacua. We show that one of them is the fivebrane giant graviton, and that its fluctuations form a single $\mathfrak{su}(2|4)$ multiplet, which reproduces the $\ell=1$ level of the spherical transverse fivebrane identified in the large $N$ limit of BMN theory. Moreover, the dynamical tensor supports new families of vacua, where the flux balances the size of the brane, and whose spectrum contains that of a fuzzy membrane. Finally, we show that BLG theory sits inside the model, both on a Coulomb branch of the massless theory and around the giant graviton of the massive one, where the two membranes are polarized into the fivebrane.}

\begin{document}
	\makeatletter
	\let\old@fpheader\@fpheader
	\renewcommand{\@fpheader}{\vspace*{-0.1cm}}
	\makeatother
    \maketitle
	\setcounter{page}{1}
	\pagenumbering{roman} 
	
	\hypersetup{
		pdftitle={Matrix model of an M5-brane in the plane-wave background},
		pdfauthor={Carmine Montella},
		pdfsubject={}
	}

\newcommand{\remove}[1]{\textcolor{red}{\sout{#1}}}

\newpage
\pagenumbering{arabic}

\clearpage
\setcounter{tocdepth}{2}
\clearpage

\section{Introduction}\label{s:intro}

The dualities among string theories have been one of the most important guides to their physics beyond perturbation theory. They relate the five superstring theories to one another, and they reveal that all of them are different limits of a single underlying eleven--dimensional theory, known as M--theory \cite{Hull:1994ys,Witten:1995ex,Townsend:1995kk}. At low energies, M--theory is described by eleven--dimensional supergravity, which, however, is non--renormalizable and only captures the dynamics well below the Planck scale. Part of its microscopic content is nevertheless encoded in the supersymmetry algebra, whose anticommutator contains, besides the momentum, a two--form and a five--form charge \cite{Cremmer:1978km, Townsend:1997wg}. These charges are carried by the M2--brane and by the M5--brane, the electric and magnetic sources of the three--form of supergravity. The two branes play in M--theory the role that the string plays in string theory, and understanding their quantum dynamics is one of the central problems of the theory.

The quantization of these extended objects has been a long--standing problem. For the membrane, the supersymmetric worldvolume action was constructed in \cite{Bergshoeff:1987cm}. In the light--cone gauge, its area--preserving diffeomorphisms can be regularized by replacing the functions on the sphere by matrices and the Poisson bracket by a commutator, so that the theory becomes a supersymmetric matrix quantum mechanics \cite{deWit:1988wri}. However, its potential has flat directions, along which the membrane can develop long and thin spikes at no cost in energy and, as a consequence, the spectrum is continuous \cite{deWit:1988xki}. For the fivebrane, the obstacles are of a different nature. The worldvolume action of a single M5--brane is known \cite{Pasti:1997gx,Bandos:1997ui,Aganagic:1997zq,Bergshoeff:1998vx,Bandos:2008fr}, but the quantization of its two--form has to take into account the self--duality of its field strength \cite{Henneaux:1988gg}. Moreover, the theory of several coincident fivebranes, the six--dimensional $(2,0)$ theory, has no known conventional Lagrangian description \cite{Witten:1995zh, Strominger:1995ac}.

\smallskip

The BFSS proposal changed this point of view \cite{Banks:1996vh}. The same matrix quantum mechanics obtained from the membrane is read as the theory of $N$ D0--branes, and it is conjectured to describe M--theory in discrete light--cone quantization, with $N$ the number of units of momentum along the light--like direction \cite{Seiberg:1997ad,Sen:1997we,Taylor:2001vb}. In this picture the continuous spectrum is not an instability, but describes multi--particle states, and the eleven--dimensional theory is conjectured to be recovered in the large $N$ limit. However, the same flat directions make the path integral difficult to control, and even the existence of the graviton as a normalizable bound state at threshold requires a careful analysis \cite{Witten:1995im,Witten:1982df,Sethi:1997pa,Yi:1997eg,Lin:2014wka}. Moreover, in BFSS theory the membranes and the fivebranes do not appear on the same footing. Membranes appear as extended configurations of the matrices, and their charges enter the supersymmetry algebra, while for the fivebrane only the longitudinal charge appears, and the transverse one is absent \cite{Banks:1996nn,Berkooz:1996is}.
Some of these obstacles are solved by the BMN model \cite{Berenstein:2002jq,Dasgupta:2002hx}, the deformation of BFSS theory on the maximally supersymmetric plane--wave backgrounds. In this model the dynamical fields acquire masses, which lift the flat directions. As a result, the spectrum becomes discrete, and the vacua are isolated and labelled by the representations of $\mathfrak{su}(2)$, whose irreducible blocks are fuzzy two--spheres, i.e.\ spherical membranes. The fivebrane is also present, but only through the vacuum structure. Indeed \cite{Maldacena:2002rb} gave evidence that, in the M-theory limit, the trivial vacuum $X=0$ describes a single spherical transverse fivebrane. More generally, they proposed that a vacuum consisting of $m$ copies of the $k$-dimensional irreducible representation of $\mathfrak{su}(2)$ describes $k$ coincident spherical transverse fivebranes in the limit $m\to\infty$ at fixed $k$, with $N=mk$. Thus, the spherical transverse fivebrane emerges as a quantum state of the BMN matrix model at large $N$ and strong coupling, but it is not described directly as a classical solution of the matrix model itself.

\smallskip

Recently, a distinct approach was proposed in \cite{Artime:2026fcr,Artime:2026mph}, where BFSS theory is extended by a five--bracket potential, and its structure constants are promoted to a dynamical tensor. The resulting $\mathcal{A}_6$ model is based on the unique simple finite--dimensional $5$--Lie algebra \cite{Ling:1993nLie,deAzcarraga:2010mr}, and it can be obtained by truncating the light--cone M5--brane on a five--sphere to its lowest harmonics, in the same spirit in which the $\mathfrak{su}(2)$ BFSS model is obtained from the spherical membrane \cite{deWit:1988wri,Artime:2026mph}. However, by construction, the model inherits many of the proprieties of $\mathfrak{su}(2)$ BFSS theory. It has many flat directions, and its round fivebrane is not a supersymmetric classical solution \cite{Artime:2026mph}. Moreover, since the model only describes the lowest spherical modes and has no large $N$ limit, the fivebrane cannot even be identified through its charge in the supersymmetry algebra either. In other words, it is still unclear how the supersymmetric fivebrane emerges from this model, and more generally whether a fivebrane can be described as a solution of a finite matrix model.

\smallskip

In this work we address these problems. We construct a maximally supersymmetric plane--wave deformation of the $\mathcal{A}_6$ model, analogous to the BMN deformation of $\mathfrak{su}(2)$ BFSS matrix theory. In the resulting massive $\mathcal{A}_6$ model, the scalar flat directions are lifted and the spherical fivebrane appears directly as a classical supersymmetric vacuum. This allows us to study the vacuum structure and the fluctuations around each vacuum explicitly.

\medskip

This work is organized as follows. In Section~\ref{sec:mass}, we first briefly review the $\mathcal A_6$ matrix model. We then construct its maximally supersymmetric plane--wave deformation. The plane--wave background induces masses for the scalars and fermions, together with two new polarization terms. We then give the equations of motion, the dynamical and kinematical supersymmetry transformations, and the supersymmetry algebra. The complete proof of supersymmetry closure is given in Appendix~\ref{app:susy}.

In Section~\ref{sec:vacua}, we classify all the classical zero-energy configurations of the constructed model. The first solution, denoted by $\mathcal S$, is a stable spherical fivebrane wrapping $\mathbf{S}^5 \subset\mathbb R^6$, with vanishing self-dual tensor. A second class of solutions extends in the complementary three--plane, where a nonzero chiral tensor field supports a three-ball $\mathbf{B}^3 \supset \mathbf{S}^2$ of finite radius. The tensor components along the occupied three directions determine its radius, while the components with legs in the remaining directions distinguish two supersymmetric families, $\mathcal R_{\mathrm I}$ and $\mathcal R_{\mathrm{II}}$. On the two extended families, we identify an underlying $\mathfrak{su}(2)$ algebra and we exhibit a limit in which the $\mathfrak{su}(2)$ BMN Lagrangian is recovered. We also find a collapsed solution for which the scalars vanish but the tensor can remain nonzero. These solutions are absent in the massless $\mathcal{A}_6$ theory.
We finally determine how the physical fluctuations organize into representations of $\mathfrak{su}(2|4)$. Around $\mathcal S$, they form a single multiplet corresponding to the $\ell=1$ harmonic of the spherical transverse fivebrane described by the large $N$ limit of BMN vacuum $X=0$ \cite{Maldacena:2002rb}. Both $\mathcal R_{\mathrm I}$ and $\mathcal R_{\mathrm{II}}$ contain the $\mathfrak{su}(2|4)$ multiplet of the BMN fuzzy two--sphere \cite{Dasgupta:2002hx}.  

In Section~\ref{sec:m5}, we first review the spherical M5--brane solution in the eleven-dimensional plane--wave \cite{Maldacena:2002rb}. Using the Hamiltonian formulation of the M5--brane \cite{Bergshoeff:1998vx} and its light-cone description \cite{Bandos:2008fr}, we extend the flat--space spherical reduction of \cite{Artime:2026mph} to the plane--wave background and show that it yields the proposed matrix model constructed in Section~\ref{sec:mass}. 
Further details are given in Appendix~\ref{app:superM5}.

Finally, in Section~\ref{sec:blg}, we show how BLG theory \cite{Bagger:2006sk,Bagger:2007jr,Gustavsson:2007vu}, reduced to $0+1$ dimensions, arises at low energies on a Coulomb branch of the massless $\mathcal A_6$ model, where two scalar expectation values reduce the five-bracket to the $\mathcal A_4$ three-bracket dynamically. In the massive theory, we relate $\mathcal S$ to the fuzzy three-sphere vacuum of massive BLG and show how the additional scalar directions fix its radius \cite{Hosomichi:2008qk,Gomis:2008cv}.
We conclude with a discussion of the results and possible directions for future work in Section \ref{sec:conclusion}.

\section{The massive \texorpdfstring{$\mathcal{A}_6$}{A6} model}\label{sec:mass}

Recently, the $\mathcal{A}_6$ extension of BFSS theory was proposed \cite{Artime:2026fcr,Artime:2026mph}, with the aim of including fivebranes and their interactions explicitly, by adding to BFSS theory a supersymmetric potential built from a five--bracket.
However, simply adding the five--bracket interactions to BFSS theory leaves various terms in the supersymmetry variation which do not cancel.
The main idea is that the extension is supersymmetric only if the BFSS structure constants $f_{abc} \to H_{abc}(t)$ become dynamical and transforms under supersymmetry as
\begin{eqn}
 \delta H_{abc}=-\frac\ii2F_{abcdef}\,(\epsilon^T\gamma_{JK}\Theta_d)X^J_eX^K_f .
\end{eqn}
The interesting part is that its variation in the kinetic terms is then cancelled by a first--order kinetic term of Chern--Simons type, provided $F$ squares to minus the identity on three--forms. The tensor $H_{abc}$ is then interpreted as the lowest harmonic of the magnetic field strength of the chiral two--form of the M5--brane.
Supersymmetry also requires the five--bracket to satisfy the fundamental identity, i.e. the generalized Jacobi identities or Filippov identities for $F$ \cite{deAzcarraga:2010mr, Ling:1993nLie}. For a positive definite metric its only simple solution is the six--dimensional one,
\begin{eqn}\label{eq:Fdef}
 F_{abcdef}=\frac{1}{3!}\,\epsilon_{abcdef},\qquad a,\dots,f=1,\dots,6,
\end{eqn}
which defines the algebra $\mathcal{A}_6$, in analogy with $\mathcal{A}_4$ in BLG theory \cite{Papadopoulos:2008sk, Gauntlett:2008uf}. 

\smallskip

In this Section we show that the theory admits a deformation which preserves maximal supersymmetry, and which reduces to the model above as particular case. We will see that this deformation is the analogue of the plane--wave deformation of $\mathfrak{su}(2)$ BMN theory for a spherical fivebrane. To be precise, in addition to the masses and to the coupling to the three--form flux, it contains a coupling to the dual six--form potential $C_6$, whose explanation will become clearer in Section~\ref{sec:m5}.


\subsection{The $\mathcal A_6$ model}\label{ss:massless}

Let us start from the BFSS matrix model \cite{Banks:1996vh}. Its dynamical fields are the transverse scalars $X^I_a$, with $I=1,\ldots,9$, and the real sixteen--component spinors $\Theta_a$, where $a$ labels a basis of the gauge algebra with structure constants $f_{abc}$ defined by $[X^I, X^J]_a= f_{abc} X^I_b X^J_c$, and $A_{ab}=f_{abc}A_c$ is the gauge connection. In this notation the BFSS Lagrangian reads\footnote{See \cite{Taylor:2001vb} for review on the subject.}
\begin{eqn}\label{LBFSS}
 L_{\rm BFSS}=\frac12(D_tX_a^I)^2+\frac\ii2\Theta_a^TD_t\Theta_a
 -\frac14\bigl(f_{abc}X_b^I X_c^J\bigr)^2
 +\frac\ii2 f_{abc}\,\Theta_a^T\gamma_I\Theta_b\,X_c^I\,.
\end{eqn}
For $\mathfrak{u}(N)$ the fields become Hermitian $N\times N$ matrices, and \eqref{LBFSS} is the quantum mechanics of $N$ D0--branes. According to \cite{Banks:1996vh}, this simple system describes M--theory in the light--cone frame, with $N$ playing the role of the momentum along the light--like direction. Membranes enter naturally in this theory through the two--bracket, since at large $N$ the $\Tr[X^I,X^J]$ is precisely the membrane charge appearing in the supersymmetry algebra \cite{Banks:1996nn}. However, nothing similar happens for the fivebrane, whose transverse charge is absent. The same commutator is also the origin of the main difficulty of the model. Indeed, the potential vanishes whenever the matrices commute, i.e.\ whenever the D0--branes are simply separated in the transverse space, and these flat directions survive quantum mechanically. The spectrum is then continuous, and even the graviton, a bound state at threshold, can be found only after a careful analysis \cite{Witten:1995im,Sethi:1997pa,Yi:1997eg}.

\smallskip

As discussed above, the $\mathcal A_6$ model is obtained by adding a five--bracket with structure constants \eqref{eq:Fdef}, and by promoting the structure constants $f_{abc}$ to a dynamical tensor $H_{abc}(t)$ with a first--order kinetic term. Its fields are therefore the following. The scalars $X_a^I$ and the spinors $\Theta_a$ carry now an internal index $a=1,\ldots,6$, which labels the spherical functions retained in \eqref{dictionary}. 
In terms of the two-- and five--brackets defined by the structure constants,
\begin{eqn}\label{brackets}
 [x,y]_a=H_{abc}\,x_by_c,\qquad
 [x_1,\ldots,x_5]_a=F_{abcdef}\,x_{1b}x_{2c}x_{3d}x_{4e}x_{5f},
\end{eqn}
the Lagrangian of the $\mathcal A_6$ model reads \cite{Artime:2026fcr,Artime:2026mph}
\begin{eqn}\label{LA6long}
 L_{\mathcal A_6}={}&\frac12\bigl(D_tX^I_a\bigr)^2+\frac12H_{abc}F_{abcdef}D_tH_{def}+\frac\ii2\Theta_a^TD_t\Theta_a\\
 &-\frac14\,[X^I,X^J]^2
 -\frac1{2\cdot5!}\,[X^I,X^J,X^K,X^L,X^M]^2\\
 &+\frac\ii2\,\Theta_a^T\gamma_I\,[\Theta,X^I]_a
 -\frac{\ii}{2\cdot4!}\,\Theta_a^T\gamma_{IJKL}\,[\Theta,X^I,X^J,X^K,X^L]_a .
\end{eqn}
The first line contains the kinetic terms, including the Chern--Simons--like term of the tensor $H(t)$. The second line is the potential, given by the squares of the two brackets, while the third line contains the corresponding Yukawa couplings. The action associated to this Lagrangian has been then proved to be invariant under the following supersymmetry transformations
\begin{eqn}\label{susy0}
 \delta X_a^I&=-\ii\epsilon^T\gamma_I\Theta_a\,, \\
 \delta\Theta_a&=\left[P_a^I\gamma_I+\frac12 H_{abc} X_b^I X_c^J \gamma_{IJ}
 +\frac1{5!} F_{abcdef} X_{b}^I X_{c}^J X_{d}^K X_{e}^L X_{f}^M \gamma_{IJKLM}\right]\epsilon,\\
 \delta H_{abc}&=-\frac\ii2 F_{abcdef}(\epsilon^T\gamma_{IJ}\Theta_d)X_e^IX_f^J\,, \\
 \delta A_{ab}&=-\ii\epsilon^T\left[H_{abc}
  -\frac1{3!}F_{abcdef}X_d^IX_e^JX_f^K\gamma_{IJK}\right]\Theta_c\,,
\end{eqn}
with $\epsilon(t)=\epsilon_0$ a constant spinor; see \cite{Artime:2026fcr, Artime:2026mph} for details.

\smallskip

Since the action contains many terms, and even more once the mass deformation is included, it is convenient to introduce a compact notation for the interactions. We collect the two interactions of the massless model as
\begin{eqn}\label{BCW}
 B_a^{IJ}&=H_{abc}X_b^IX_c^J,\qquad
 C_a^{IJKLM}=F_{abcdef}X_b^IX_c^JX_d^KX_e^LX_f^M,\\
 (W_H)_{ab}&=H_{abc}X_c^I\gamma_I,\qquad
 (W_F)_{ab}=\frac1{4!}F_{abcdef}X_c^IX_d^JX_e^KX_f^L\gamma_{IJKL},
 \qquad W_\pm=W_H\pm W_F .
\end{eqn}
The tensor $B$ describes the interaction between the three--form to two scalar vectors, while $C$ measures the five--volume spanned by five of them, or the five scalars self--interactions. Their fermionic counterparts are $W_H$ and $W_F$, which are antisymmetric matrices on the combined internal and spinor space. In this work we use real symmetric Clifford matrices, with
\begin{eqn}\label{cliff}
 \{\gamma_I,\gamma_J\}=2\delta_{IJ},\qquad \gamma_I^T=\gamma_I,\qquad
 \gamma_1\cdots\gamma_9=1,\\
 G=\gamma_{123}=-\gamma_{456789},\qquad G^T=-G,\qquad G^2=-1,
\end{eqn}
and the covariant derivatives act on every internal index,
\begin{eqn}\label{covariantderivatives}
 D_tX_a^I&=\dot X_a^I-A_{ab}X_b^I,\qquad
 D_t\Theta_a=\dot\Theta_a-A_{ab}\Theta_b,\\
 (D_tH)_{abc}&=\dot H_{abc}-A_{ad}H_{dbc}-A_{bd}H_{adc}-A_{cd}H_{abd}.
\end{eqn}
With these definitions, the massless action of \cite{Artime:2026fcr,Artime:2026mph} takes the form
\begin{eqn}\label{Lzero}
 L_0={}&\frac12(D_tX)^2+\frac12H D_tH
 +\frac\ii2\Theta^TD_t\Theta\\
 &-\frac{1}{2\cdot 2!}B^2-\frac1{2 \cdot 5!}C^2
 +\frac\ii2\Theta^T W_-\Theta \,,
\end{eqn}
where we omitted the indices between $H$'s indicating the contraction via the tensor $F$. As one can already see from the above Lagrangian, this massless model has the same issue as BFSS theory, in an even stronger form. Indeed, its potential is a sum of two squares, and it vanishes whenever both brackets vanish. Since $H$ is dynamical, the two--bracket can be switched off simply by setting $H=0$, while the five--bracket vanishes whenever the scalars span fewer than five internal directions. Hence every such configuration has zero energy and it also satisfies the Gauss' law at zero momentum. This gives a large family of flat directions, which make the spectrum and the path integral difficult to control. Indeed, the round fivebrane is not a static supersymmetric solution of the massless model, but requires possibly a rotating flux which breaks supersymmetry \cite{Artime:2026mph}. In the next Section we show that both problems are solved by the allowing the most general supersymmetric deformation, which lifts the flat directions of the scalars and turns the round fivebrane into a supersymmetric vacuum. 

\medskip

To conclude this section, we record that the tensor kinetic term is first order in time, as expected for the phase-space description of a self-dual field \cite{Henneaux:1988gg,Bergshoeff:1998vx,Bandos:2008fr}. Consequently, its twenty components cannot be assigned all independent momenta. In Appendix~\ref{app:details} we compute the Dirac brackets
\begin{eqn}\label{dirac}
 \{X_a^I,P_b^J\}_D=\delta_{ab}\delta^{IJ},\qquad
 \{H_p,H_q\}_D=-\frac16F_{pq},\qquad
 \{\Theta_{a\lambda},\Theta_{b\sigma}\}_D=-\ii\delta_{ab}\delta_{\lambda\sigma}\,,
\end{eqn}
where we labelled the triples by $p$ and $q$. As confirmed by the Dirac brackets every two triple components form only one canonical pair, leaving only ten tensor pairs in total.
Finally, varying the gauge connection in \eqref{Lzero} imposes the internal angular--momentum constraint or Gauss' law \cite{Artime:2026fcr, Artime:2026mph},
\begin{eqn}\label{gauss}
 \Ga_{ab}&\equiv X_{[a}^IP_{b]}^I-\frac\ii2\Theta_a^T\Theta_b-\frac32K_{ab}=0,\\
 K_{ab}&\equiv -(FH)_{[a|cd|}H_{b]cd},\qquad
 (FH)_{abc}\equiv \frac16\epsilon_{abcdef}H_{def} .
\end{eqn}

The three terms in $\Ga$ are the common scalar orbital contribution, the fermion spin contribution and the tensor moment map. Hence, the term $-\Lambda_{ab}\Ga_{ab}$ generates the rotation $\delta X_a=\Lambda_{ab}X_b$ through \eqref{dirac}. 

\subsection{The plane--wave deformation}\label{ss:massdeformation}


Besides flat space, M--theory has three other maximally supersymmetric backgrounds, $AdS_4\times S^7$, $AdS_7\times S^4$ and the eleven dimensional pp--wave background \cite{Kowalski-Glikman:1984qtj,Figueroa-OFarrill:2002ecq}. The latter can be obtained from both of the former as a Penrose limit \cite{Blau:2002dy}. In this background the transverse directions acquire masses, and the four--form flux couples to the branes through a polarization term, which makes them expand into spheres stabilized by the flux, as in the Myers effect \cite{Myers:1999ps}. This is the mechanism behind the BMN model, where the flat directions are lifted, the vacua are now isolated fuzzy two--spheres, differently from the massless counterpart \cite{Dasgupta:2002hx,Maldacena:2002rb, Berenstein:2002jq}. Since the $\mathcal{A}_6$ model encodes the lowest--harmonic truncation of a spherical M5--brane, it is natural to place it in the same background. However, differently from BFSS theory, the $\mathcal{A}_6$ model has no large $N$ limit. Hence, the fivebrane can be identified neither through the supersymmetry algebra nor through a supersymmetric solution of the massless model. 

A natural way to obtain the fivebrane as a classical supersymmetric solution is therefore to place the model in the pp--wave background, allowing for the most general maximally supersymmetric deformation, which we call the massive $\mathcal{A}_6$ model.

\medskip

To start, the plane--wave metric \eqref{ppwave} distinguishes a three--plane and a six-plane, so we split $X=(Y,Z)$ with $i=1,2,3$ and $m=4,\ldots,9$. Already from the splitting we can see that the background three-form $C_3$ suggests a coupling involving $H$ and three $Y$'s, while its dual six-form suggests a coupling involving six $Z$'s. 
To simplify the notation, we introduce
\begin{eqn}\label{shiftedtensors}
 (E_2)_a^{ij}=\epsilon^{ijk}Y_a^k,\qquad
 (E_5)_a^{m_1\ldots m_5}=\epsilon^{n m_1\ldots m_5}Z_a^n,
\end{eqn}
with all other target components set to zero, and propose the shifted potential
\begin{eqn}\label{Vsquares}
 V_{\alpha,\beta}&=\frac14(B-\alpha E_2)^2
                  +\frac1{2\,5!}(C-\beta E_5)^2 \\
                   &=V_0+\frac{\alpha^2}{2}Y^2+\frac{\beta^2}{2}Z^2
 -\frac\alpha2\epsilon^{ijk}H_{abc}Y_a^iY_b^jY_c^k
 -\frac\beta{5!}\epsilon^{n m_1\dots m_5} \epsilon_{abcdef} Z_a^n \, Z_b^{m_1} Z_c^{m_2}Z_d^{m_3}Z_e^{m_4}Z_f^{m_5} \\
                  &= V_0+\frac{\alpha^2}{2}Y^2+\frac{\beta^2}{2}Z^2 - \frac{\alpha}{2} E_2 \cdot B - \frac{\beta}{5!} E_5 \cdot C
\end{eqn}

The mass coefficients follow from $E_2^2=2Y^2$ and $E_5^2=5!\,Z^2$. Moreover, contracting the two six-dimensional epsilon tensors gives
\begin{eqn}\label{detflux}
 \frac1{5!}(E_5)_a^{m_1\ldots m_5}C_a^{m_1\ldots m_5}=\det Z .
\end{eqn}
Hence, the six-form coupling therefore measures the (oriented) volume of the scalar frame $Z$, whereas the three-form coupling measures the scalar three-frame contracted with $H$. 

\medskip

However, the most pressing question is whether this general deformation admits a maximally supersymmetric matrix model realization as for the BMN theory \cite{Berenstein:2002jq,Dasgupta:2002hx,Kim:2006wg}. The answer of this question is affirmative since, as we show below, supersymmetry fixes the relation between the two shifts. To prove this statement, as for the BMN matrix theory, we add a fermion mass and a scalar linear term to the massless fermion transformation,\footnote{The signs of $c_i$ and $c_m$ follow by expressing the shifts in \eqref{shiftedtensors} through the Clifford dualities in \eqref{cliff}. In general we can consider different coefficients, and the supersymmetry closure of the theory will give an equivalent result (see eq. \eqref{susy:km})}
\begin{eqn}\label{massansatz}
 \Delta L_F=-\frac{\ii m_F}{2}\Theta_a^TG\Theta_a,\qquad
 \delta_c\Theta_a=c_I X_a^I\gamma_IG\epsilon,\qquad
 c_i=-\alpha,\quad c_m=\beta .
\end{eqn}
Since the variation of the new terms can be cancelled by the term proportional to $\dot\epsilon$ in the variation of the massless action, we allow $\epsilon$ to depend on time. At this stage $\alpha$, $\beta$ and $m_F$ are independent, and we collect the details of the closure in Appendix~\ref{app:susy}.

\medskip

In short, the first condition comes from the terms containing one scalar velocity and one fermion. Their cancellation determines both the allowed time dependence of $\epsilon(t)$ and the fermion mass,
\begin{eqn}\label{noetherlinear}
 \dot\epsilon=kG\epsilon,\qquad
 k=\frac{\alpha-\beta}{2},\qquad m_F=-\frac{\alpha+\beta}{2}.
\end{eqn}
Finally, the interactions containing $HX^2\Theta$ and $FX^5\Theta$ both require 
\begin{eqn}
    \alpha=2\beta\,.
\end{eqn}
With the usual plane--wave parameter $\mu>0$, the result is
\begin{equation}\label{massrelations}
 \alpha=\frac{\mu}{3},\qquad
 \beta=\frac{\mu}{6},\qquad
 k=\frac{\mu}{12},\qquad
 m_F=-\frac{\mu}{4}.
\end{equation}

Thus the relative masses, polarization couplings and rotation rate of the supersymmetry parameter are fixed by one scale $\mu$. In the massless limit all these coefficients vanish, and a constant supersymmetry parameter is recovered. The completed action is therefore
\begin{eqn}\label{Lmass}
 L_\mu={}&\frac12(D_tX)^2+\frac12HFD_tH+\frac\ii2\Theta^TD_t\Theta
 -V_{\alpha,\beta}+\frac\ii2\Theta^TW_-\Theta+\frac{\ii\mu}{8}\Theta^TG\Theta,\\
 V_{\alpha,\beta}&=V_0+\frac{\alpha^2}{2}Y^2+\frac{\beta^2}{2}Z^2
 -\frac\alpha2\epsilon^{ijk}H_{abc}Y_a^iY_b^jY_c^k
 -\beta \det Z\,,
\end{eqn}
where again $\alpha=\frac\mu3,\quad\beta=\frac\mu6$.
Finally, although this method gives the finite maximally supersymmetric theory already by construction, it does not specify a precise spacetime interpretation. We postpone this analysis in Section~\ref{sec:m5} where we will provide an independent geometric derivation of the same theory from the light--cone plane--wave M5 action in its spherical sector.

\subsubsection*{Supersymmetry transformations and equations of motion}
Substituting \eqref{massrelations} and \eqref{massansatz} into \eqref{susy0} we obtain the following supersymmetry transformations
\begin{eqn}\label{susy}
 \delta X_a^I&=-\ii\epsilon^T\gamma_I\Theta_a,\\
 \delta\Theta_a&=\left[P_a^I\gamma_I+\frac12(B-\alpha E_2)_a^{IJ}\gamma_{IJ}
 +\frac1{5!}(C-\beta E_5)_a^{IJKLM}\gamma_{IJKLM}\right]\epsilon,\\
 \delta H_{abc}&=-\frac\ii2 F_{abcdef}(\epsilon^T\gamma_{IJ}\Theta_d)X_e^IX_f^J,\\
 \delta A_{ab}&=-\ii\epsilon^T\left[H_{abc}
  -\frac1{3!}F_{abcdef}X_d^IX_e^JX_f^K\gamma_{IJK}\right]\Theta_c,\\
 \epsilon(t)&=e^{\mu tG/12}\epsilon_0 .
\end{eqn}

The constant spinor $\epsilon_0$ has sixteen independent real components. Note that the shifted brackets appear directly and exclusively in $\delta\Theta$, so a static bosonic configuration for which both shifts vanish preserves all sixteen transformations, if the Gauss' law is also satisfied.

\smallskip

As in the massless construction, we can include
sixteen additional kinematical supersymmetries by adjoining the
constant spherical harmonic. We include its scalar and fermion
coefficients by $X_0^I=(Y_0^i,Z_0^m)$ and $\Theta_0$, and set
$A_{0a}=H_{0ab}=F_{0abcde}=0$ as in \cite{Artime:2026fcr,Artime:2026mph}. These fields form a decoupled
center--of--mass multiplet, whose plane--wave Lagrangian is
\begin{equation}
\begin{aligned}
L_{\mathrm{cm}}={}&
\frac12\dot X_0^I\dot X_0^I
-\frac12\left(\frac{\mu}{3}\right)^2Y_0^iY_0^i
-\frac12\left(\frac{\mu}{6}\right)^2Z_0^mZ_0^m\\
&+\frac{i}{2}\Theta_0^T\dot\Theta_0
+\frac{i\mu}{8}\Theta_0^TG\Theta_0 .
\end{aligned}
\end{equation}
This is the free plane--wave multiplet familiar from the abelian
sector of the BMN model \cite{Berenstein:2002jq,Dasgupta:2002hx}. The dynamical transformations
extend to it by setting the bracket interactions to zero. In
addition, the full action is invariant, up to a total derivative,
under
\begin{equation}\label{eq:kinsusy}
\delta_\eta\Theta_0=\eta(t),\qquad
\eta(t)=e^{-\mu tG/4}\eta_0 ,
\end{equation}
with all other fields unchanged and $\eta_0$ an arbitrary constant
sixteen-component Majorana spinor. Together with the sixteen
dynamical supersymmetries, these transformations give thirty-two
supersymmetries. We leave this decoupled sector implicit in the
following analysis of the interacting massive $\mathcal A_6$ model and we give more information in Appendix~\ref{app:superM5}.

\medskip


We now construct the supersymmetry charges. Since the parameter $\epsilon(t)$ depends on time, the charge $Q$ is not conserved by itself. However, as in the massless theory, the combination $S_\epsilon=-\ii\epsilon(t)^TQ$ is conserved, and it generates the transformations \eqref{susy} through the Dirac brackets \eqref{dirac}. Requiring this fixes the charge to be

\begin{eqn}\label{Q}
 Q=\sum_a\left[P_a^I\gamma_I-\frac12B_a^{IJ}\gamma_{IJ}
 +\frac1{5!}C_a^{IJKLM}\gamma_{IJKLM}
 +\alpha Y_a^i\gamma_iG+\beta Z_a^m\gamma_mG\right]\Theta_a .
\end{eqn}
Notice that the sign of the two--gamma term differs from the one in $\delta\Theta$, since $\gamma_{IJ}$ is antisymmetric under transposition. For the same reason, the target angular momentum must include the spin of the fermions,
\begin{eqn}\label{targetJ}
 \mathcal J_{IJ}=\sum_a(X_a^IP_a^J-X_a^JP_a^I)
                  -\frac\ii4\sum_a\Theta_a^T\gamma_{IJ}\Theta_a .
\end{eqn}
Finally, to compute the algebra we also need the Hamiltonian. The Legendre transform of \eqref{Lmass}, after eliminating the momentum constraints of the tensor and of the fermions, trivially gives
\begin{eqn}\label{Hamiltonian}
 \Ham=\frac12P^2+V_{\alpha,\beta}
       -\frac\ii2\Theta^TW_-\Theta-\frac{\ii\mu}{8}\Theta^TG\Theta .
\end{eqn}
With these ingredients, the brackets of the charges can be evaluated on the (constrained) phase space. As shown in Appendix~\ref{app:susy}, they close into
\begin{eqn}\label{algebra}
 \{Q_\lambda,Q_\sigma\}_D={}&-2\ii\delta_{\lambda\sigma}\Ham
 +2\ii\Ga_{ab}(W_+)_{ab,\lambda\sigma}\\
 &-\ii\alpha(\gamma_{ij}G)_{\lambda\sigma}\mathcal J_{ij}
 +\ii\beta(\gamma_{mn}G)_{\lambda\sigma}\mathcal J_{mn} .
\end{eqn}
In this formula each term has a clear physical interpretation. The first term generates time evolution, and the term containing $\Ga$ generates an internal gauge rotation. The remaining terms generate rotations of the physical target coordinates in the two planes, which are generically non--zero even after imposing the Gauss' law. This is the same structure as in BMN theory \cite{Dasgupta:2002hx}, with the coefficients $\alpha=\frac\mu3$ and $\beta=\frac\mu6$, which here are fixed by supersymmetry.
They disappear only as $\mu\to0$, recovering the massless algebra of \cite{Artime:2026fcr,Artime:2026mph}.

For later use, Hamilton's equations following from \eqref{dirac} and \eqref{Hamiltonian} are
\begin{eqn}\label{EOM}
 D_tX_a^I&=P_a^I\,,\qquad 
 D_tP_a^I=-\partial_{X_a^I}V_{\alpha,\beta}
                 +\frac\ii2\Theta^T\partial_{X_a^I}W_-\Theta\\
 D_t h_p&=-\frac16F_{pu}\left(\partial_{h_u}V_{\alpha,\beta}
                 -\frac\ii2\Theta^T\partial_{h_u}W_H\Theta\right) \\  
 D_t\Theta&+\mathcal W\Theta=0,\qquad \Ga_{ab}=0 .
\end{eqn}

where $\mathcal W=W_-+\mu G/4$. The tensor evolves by applying $F_{pu}$ to the gradient of its Hamiltonian, which is the counterpart of the two equations for a canonical coordinate and its momentum. 


\section{Supersymmetric static solutions}\label{sec:vacua}

We now classify the bosonic zero--energy configurations of the massive model. Since the potential is a sum of squares, a configuration has zero energy if and only if both shifted brackets vanish (and the Gauss' law is satisfied). By \eqref{susy}, such a configuration is static and preserves all sixteen supersymmetries. The main new feature is that the structure constants of the two--bracket are replaced by the dynamical tensor $H_{abc}(t)$, whose size can dynamically compensate the size of the scalars. As a result, besides an isolated spherical solution, we find continuous families of supersymmetric configurations supported by the flux, and a cone of pure flux.\footnote{Throughout this section, with a slight abuse of notation, we call vacuum any static bosonic zero of the classical Hamiltonian. Whether such a configuration supports a quantum ground state is a different question, which requires a separate analysis and is postponed to future work.}
In Section~\ref{ss:ranks} we derive the complete classification, and in Section~\ref{ss:tensoralgebra} we discuss the algebra induced by the chiral tensor and the associated BMN limit. In Sections~\ref{ss:sphere} and \ref{ys:section} we study in more detail the spherical vacuum and the flux--supported families, together with their fluctuations.

\subsection{The zero--energy configurations}\label{ss:ranks}

The bosonic Hamiltonian is a sum of non--negative terms, so it vanishes only if each of them does. This happens when the momenta vanish and both brackets are equal to their shifts,
\begin{eqn}\label{vacconditions}
 P=0,\qquad B=\alpha E_2,\qquad C=\beta E_5,\qquad K=0 \,,
\end{eqn}
where the last condition is what remains of the Gauss' law \eqref{gauss} once the momenta are set to zero. At a zero of the two squares every first derivative of the potential vanishes, also the one with respect to $H$, so Hamilton's equations leave the configuration at rest. At the same time, the shifted brackets are exactly the combinations entering $\delta\Theta$ in \eqref{susy}, which therefore vanishes for all sixteen dynamical supersymmetries. We now start classifying all such solutions.


\paragraph{Fivebrane giant graviton.} The five-bracket equation sharply restricts the six-plane scalars. Regard $Z_a^m$ as a $6\times6$ matrix, and using $F=\epsilon/6$, we find that contracting the all-six-plane part of $C=\beta E_5$ with $\epsilon^{n m_1\ldots m_5}/5!$ the result is
\begin{eqn}\label{cofactorvac}
 \cof Z=6\beta Z=\mu Z,\qquad Z^T\cof Z=(\det Z)\mathbf1_6 .
\end{eqn}
This reduces to a simple problem of linear algebra. The cofactor entries are five-by-five minors, so they vanish at ranks one through four and form just a rank-one matrix at rank five. 
However, if the matrix $Z$ is of rank five, and its associated cofactor matrix is of rank one, then \eqref{cofactorvac} implies a relation between two matrices of different ranks. Hence no nonzero $Z$ of rank five or lower solves the equation.

\smallskip

Consequently $Z$ is either zero or a rank--six matrix. In the latter case the second equation fixes all singular values to 
\begin{eqn}
    R_0= \mu^{1/4}
\end{eqn}
and fixes the orientation, giving $Z=\mu^{1/4}\,O_6$ with $O_6\in SO(6)$.
The remaining bracket equations then determine the other fields. The components of $C$ with one $Y$ and four $Z$'s force $Y=0$, because the $Z$ columns span the internal space. The equation $B^{mn}=0$ similarly forces $H=0$. Finally, due to the $SO(6)$ invariance, after an internal rotation there is only one orbit with nonzero $Z$,
\begin{eqn}\label{spherevac}
 \mathcal S:\qquad Y=0,\quad Z=\mu^{1/4}\mathbf1_6,\quad H=0 .
\end{eqn}
The solutions studied above acquire a precise geometric meaning through the (extended) spherical dictionary of \cite{Artime:2026mph}. The map $Z^m=R_0u_{m-3}$ embeds a regular $\mathbf{S}^5 $ in the six-plane $\mathbb{R}^6$, with radius supported by the background six--form. This exactly reminds to a possible supersymmetric static spherical fivebrane classical solution, equivalent to the fuzzy two--sphere in BMN, since both are giant gravitons. We will extend this dictionary in Section~\ref{sec:m5}, and we will study the small fluctuations around it in Section~\ref{ss:sphere}.

\medskip

\paragraph{The two flux--supported solutions.} As anticipated at the beginning of this Section, the presence of a time--dependent tensor $H_{abc}(t)$ makes this solution space not trivial. Visually speaking, the reason is that we may consider a configuration in $\mathbb{R}^3$ which does not collapse due to the three--tensor internal fluxes.
Indeed, when $Z=0$ the five-bracket vanishes identically, and the scalar equation reduces to $H_{abc}Y_b^iY_c^j=\alpha\epsilon^{ijk}Y_a^k$. However, this relation is the three--dimensional analogous of \eqref{cofactorvac}. This means we can similarly proceed as before. The main difference is that the physics of this solution must not be characterized only by the mass shift $\mu$, but also on the value, and in particular on its tensor structure components, of the $H$--tensor. 

\smallskip

If $Y$ has rank three we can then consider again $Y_i{}^j=r_i\delta_i{}^j$ as solution (and complementary internal indices $A=4,5,6$). With $h=H_{123}$ the equations are then
\begin{eqn}
 hr_2r_3=\alpha r_1,\qquad hr_3r_1=\alpha r_2,\qquad
 hr_1r_2=\alpha r_3\,,
\end{eqn}
which imply $r_1=r_2=r_3=r$ and $h=\alpha/r$ (the complementary equations imply $H_{Aij}=0$). Finally, since $Y^TY=r^2\mathbf1$, an internal $SO(6)$ transformation alone can always represents the vector in the form $Y=(r\mathbf\mathbf{1}_3,0)^T$. Summarizing, this solution is due to
\begin{eqn}
    \mathcal{R}:\quad  Z = 0\,, \quad Y=(r\mathbf\mathbf{1}_3,0)^T \,, \quad  h = H_{123} = \frac{\alpha}{r}
\end{eqn}
It is compelling to see that the increasing the scalar size can be compensated by decreasing a tensor component, leaving a continuous family of solutions at classical level despite the nonzero masses. In other words, one of the characteristic of $\mathcal{R}$ is indeed that the balance fixes $hr=\alpha$, rather than fixing $r$ itself! 

\medskip

Interestingly this solution does not imply a full vanish solution for the 3--forms $H_{ABC}$ and $H_{iAB}$. In this representative the most general tensor allowed by the shifted two bracket equations is in general
\begin{eqn}\label{rankthreeH}
 H_{123}=h=\frac\alpha r,\quad H_{Aij}=0,\quad
 H_{ABC}=k\epsilon_{ABC},\quad H_{iAB}=\epsilon_{ABC}M_{iC} .
\end{eqn}

The matrix $M$ is in principle not specified by $B = \alpha E_2$. However, the Gauss' law depends on $H$. This means that one triple has no trivial background source for $H$ and the Gauss' law (see eq. \eqref{gauss}) may provide non--trivial information about the generic matrix $M$. If written in terms of the $M_{iA}$ it gives
\begin{eqn}\label{eq:K1}
 K_{ij}&=K_{AB}=0\,,\\
 K_{iA}&=2\bigl((\cof M)_{iA}-hM_{iA}\bigr) = 0.
\end{eqn}
So, again, $\cof M=hM$, and the positive $h$ fixes $M=h\,O_3$. In this case, however, we can also consider $h=0$ as non--trivial solution. Using the residual $SO(3)$ symmetry, we therefore obtain two families,
\begin{eqn}\label{rankthreebranches}
 &\mathcal R_{\mathrm I}:\quad M=0,\qquad \\
 &\mathcal R_{\mathrm {II}}:\quad M=h\mathbf\mathbf{1}_3,
 \qquad r>0,\quad k\in\mathbb R .
\end{eqn}

In the first family the complementary frame can still rotate, giving an $SO(3)$ stabilizer. In the second, the mixed tensor $M=h\mathbf\mathbf{1}_3$ locks it to the occupied frame, so the stabilizer is trivial. This distinction becomes clearer in terms of algebra induced by the tensor in Section \ref{ss:tensoralgebra}. 


\paragraph{The cone of pure flux.}
The simplest solution for which the scalar potential vanishes identically is then $X=0$. However, the tensor can still be nonzero, so that the collapsed configuration carries a residual flux. This is the M--theory analogue of a D2--brane wrapped on a collapsing two--cycle, where the internal flux, i.e.\ the dissolved D0--charge, survives the collapse of the cycle and keeps its physical effects. Indeed, we propose to interpret it as a fivebrane shrank to a point in the $\mathcal{R}$ vacuum which still supports internal fluxes on its worldvolume (see Section \ref{ys:section}). Since on the M5--brane the self--dual three--form carries membrane charge, these fluxes describe membrane charge stored on a fivebrane of zero size. Interestingly, the energy of this configuration still vanishes because the flux couples to the scalars only through the size of the brane, so once the fivebrane has collapsed the flux no longer costs energy. 

\smallskip

Studying the Gauss' law $K(H)=0$ (see again eq. \eqref{gauss}) one can show that the only non--zero components are given by
\begin{eqn}\label{tensorNFform}
 H=a\,e^{123}+b\,e^{456},\qquad a,b\in\mathbb R\,,
\end{eqn}
where $e^{abc}=e^a\wedge e^b\wedge e^c$ in the usual (oriented orthonormal) internal basis.
However, different values of $(a,b)$ can be related by a gauge transformation. Indeed, exchanging the two triples is not an $SO(6)$ rotation, unless we also flip the sign of one of them. This gives $e^{123}\to e^{456}$ and $e^{456}\to -e^{123}$, so that $(a,b)\to(-b,a)$, which is a rotation by a quarter of a turn,
\begin{eqn}\label{conequotiont}
 (a,b)\sim(-b,a)\sim(-a,-b)\sim(b,-a),\qquad
 K^{-1}(0)/SO(6)=\mathbb R^2/\mathbb Z_4 .
\end{eqn}
Hence, every nonzero tensor has stabilizer $SO(3)\times SO(3)$, which does not switch the two 3--plane, whereas at the apex $H=0$ the full $SO(6)$ is restored.


\subsection{The tensor algebra and the BMN limit}\label{ss:tensoralgebra}

To understand the geometric meaning of these solutions, we translate the matrix fields into an embedding of the fivebrane, using the dictionary of Section~\ref{sec:m5}. The internal index $a$ labels the six lowest harmonics of the five--sphere, which are its embedding coordinates $u_a(\sigma)$, so that
\begin{eqn}\label{dictionary}
 X^I(\sigma)=X^I_a\,u_a(\sigma),\qquad u_au_a=1 .
\end{eqn}
We discuss $\mathcal{S}$ in the next section, and consider here the rank--three families. Since only $Y^i_a=r\,\delta^i_a$ is nonzero, the embedding involves only $u_1,u_2,u_3$ as
\begin{eqn}\label{rank3map}
 X^i=r u_i,\qquad X^m=0,\qquad
 (X^1)^2 + (X^2)^2+(X^3)^2\le r^2\,.
\end{eqn}
Interestingly enough, the image fills a three-ball and not a 2--sphere $\mathbf{S}^2$. To clarify the geometric picture it is very productive to consider the following parametrization of the unit sphere:
\begin{eqn}\label{joinpicture}
 u=(\cos\chi\,n,\sin\chi\,m),\qquad n,m\in \mathbf{S}^2,\quad 0\leq\chi\leq\pi/2,\\
 d\Omega_5^2=d\chi^2+\cos^2\chi\,d\Omega_{2,n}^2
                         +\sin^2\chi\,d\Omega_{2,m}^2 .
\end{eqn}
Thus a point of the unit sphere $\mathbf{S}^5\subset\mathbb{R}^6$ is specified by a direction $n$ in the first three--plane, a direction $m$ in the second one, and an angle $\chi$ which controls the relative size of the two two--spheres. At a generic value of $\chi$ we see $\mathbf{S}^2\times\mathbf{S}^2$, while at $\chi=0$ only the first two--sphere survives, with unit radius, and at $\chi=\pi/2$ only the second one. These shrinking spheres are not boundaries or singularities, but the usual degeneration of polar coordinates, so all the directions $m$ are identified at $\chi=0$, and all the directions $n$ at $\chi=\pi/2$.\footnote{One could also extend $\chi$ to $[0,2\pi)$. In this case there are four equivalent sectors, related by the changes of orientation $n\to-n$ and $m\to-m$.}

This explains the shape of $\mathcal{R}$. In these coordinates the embedding \eqref{rank3map} reads $X^i=r\cos\chi\,n^i$, so it depends only on $n$ and $\chi$, while the second two--sphere is completely invisible. For fixed $\chi$ we therefore obtain a two--sphere of radius $r\cos\chi$, and as $\chi$ goes from $0$ to $\pi/2$ this radius decreases from $r$ to zero. The image is thus a three--ball, foliated by concentric two--spheres, rather than a single two--sphere. In other words, the fivebrane is collapsed onto the three--plane, and its five--dimensional worldvolume is projected onto a three--dimensional region. This is the main difference with respect to a membrane, whose two--dimensional worldvolume would be mapped onto a single two--sphere.

There is nevertheless a spherical membrane algebra within this geometry. Evaluating the tensor $B^{ij}$ (see eq. \eqref{tensorbivector}) on the occupied scalar coordinates gives
\begin{eqn}\label{membranebracket}
 B^{ij}= [X^i, X^j]_H = -6hr\,\epsilon^{ijk}X^k\,.
\end{eqn}
This is the linear $\mathfrak{su}(2)$ Poisson bracket, whose Casimir $\rho^2=\sum_i(X^i)^2$ labels spherical (symplectic) leaves. A leaf at fixed $\rho$ has the algebra used for a spherical membrane, but the present embedding includes every $0\leq\rho\leq r$. Thus the bracket explains only part of the classical geometry of $\mathcal{R}$ without turning the whole configuration into a single spherical 2--brane.

\medskip

In this sense, the tensor defines a bracket on the internal basis through $[e_a,e_b]_H=H_{abc}e_c$. For branch I, it gives
\begin{eqn}\label{fluxbranchI}
 [e_i,e_j]_H=h\epsilon_{ijk}e_k,\qquad
 [f_i,f_j]_H=k\epsilon_{ijk}f_k,\qquad [e_i,f_j]_H=0,
 \qquad f_i=e_{i+3} .
\end{eqn}
For $k\ne0$ the two triples generate commuting $\mathfrak{su}(2)$ factors, while at $k=0$ the second triple is abelian.\footnote{The commuting brackets here concern the basis vectors of the Lie algebra induced by $H$. They are different from the canonical brackets between the coefficients of $H$, so in this sense $[e_i,f_j]_H=0$ is compatible with $\{H_{123},H_{456}\}_D=-1/6$ (see eq. \eqref{dirac}).}
For branch II, the mixed components of $H$ instead give
\begin{eqn}\label{fluxbranchII}
 [e_i,e_j]_H&=h\epsilon_{ijk}e_k,\qquad
 [e_i,f_j]_H=h\epsilon_{ijk}f_k,\\
 [f_i,f_j]_H&=\epsilon_{ijk}(he_k+kf_k) .
\end{eqn}

Differently from branch I, the mixed components of the tensor couple the two triples, and the bracket of two $f_i$ produces also the $e_i$. Hence the algebra is no longer manifestly a direct sum, and it is natural to ask whether it still splits into two independent $\mathfrak{su}(2)$ factors. To answer this, we look for its ideals, i.e.\ the subalgebras that are preserved by the bracket with any element of the full algebra. Since the two triples transform in the same way, we consider the combinations $v_i=(e_i+a f_i)/\sqrt{1+a^2}$, and we require that their brackets close on the same combination, $[v_i,v_j]_H\propto\epsilon_{ijk}v_k$. Using \eqref{fluxbranchII} this gives two roots define the new basis,
\begin{eqn}\label{fluxideals}
 a_\pm=\frac{k\pm\sqrt{k^2+4h^2}}{2h},\qquad
 v_i^\pm=\frac{e_i+a_\pm f_i}{\sqrt{1+a_\pm^2}} .
\end{eqn}

In this basis,
\begin{eqn}\label{idealbrackets}
 [v_i^+,v_j^-]_H=0,\qquad
 [v_i^\pm,v_j^\pm]_H=h\sqrt{1+a_\pm^2}\,\epsilon_{ijk}v_k^\pm .
\end{eqn}
Hence, for every real $k$, the algebra is again $\mathfrak{su}(2)\oplus\mathfrak{su}(2)$. The difference with respect to branch I lies in how the scalars sit inside it. On branch I the occupied vectors $e_i$ generate one of the two factors, while the other one acts only on the unoccupied triple. On branch II, instead, each $e_i$ has a nonzero component along both $v_i^+$ and $v_i^-$, so the $\mathfrak{su}(2)$ generated by the scalars is embedded diagonally in the two factors. This is the algebraic reason why the stabilizer of branch II is trivial: a rotation of the occupied triple must be accompanied by the same rotation of the unoccupied one.

\paragraph{The large-flux limit and BMN dynamics}
Due to the natural appearance of $\mathfrak{su}(2)$ sub--algebra from these models, it is natural to wonder whether there is a direct limit which would lead to a $\mathfrak{su}(2)$ BMN sub--sector of the entire theory, similar to \cite{Artime:2026mph}. It is then natural to consider a limit where the kinetic terms, the 2--commutator, and the masses of the dynamical fields scale with the same rate while the five-bracket potential, its Yukawa coupling and the six-form term vanish. We can then take a fixed tensor $f$ satisfying $K(f)=0$ and scale
\begin{eqn}\label{largefluxscaling}
 H=\lambda f+\lambda^{-3}\eta,\qquad
 X=\lambda^{-1}x,\qquad \Theta=\lambda^{-1}\theta,
 \qquad \lambda\to\infty,
\end{eqn}
with $\mu$ fixed and with the gauge connection preserving $f$. In the rescaled action $\lambda^2L_\mu$ the limit returns
\begin{eqn}\label{massivefluxlimit}
 \lambda^2L_\mu \longrightarrow{} \quad &L_{BMN} = L_{\mathrm{YM}}
 -\frac{\alpha^2}{2}\sum_i|x^i|^2
 -\frac{\beta^2}{2}\sum_m|x^m|^2\\
 &+\frac\alpha2\epsilon^{ijk}f_{abc}x_a^ix_b^jx_c^k
 +\frac{\ii\mu}{8}\theta^TG\theta .
\end{eqn}
Here $L_{\mathrm{YM}}$ is \eqref{Lzero} with $H=f$ fixed, the tensor kinetic term omitted and the five-bracket terms removed, namely the BFSS Lagrangian \eqref{LBFSS}
\begin{eqn}
    L_{YM}={}&\frac12(D_tX)^2
 +\frac\ii2\Theta^TD_t\Theta
 -\frac{1}{2\cdot 2!}B_f^2
 +\frac\ii2\Theta^T W_H\Theta .
\end{eqn}
Equation~\eqref{massivefluxlimit} has the BMN masses and cubic coupling \cite{Berenstein:2002jq,Dasgupta:2002hx}. Its generic semisimple algebra consists of two $\mathfrak{su}(2)$ factors, with abelian directions when one coefficient in the normal form of $f$ vanishes. This identifies a formal $\mathfrak{su}(2)$ BMN action limit of the massive $\mathcal{A}_6$ model.

\subsection{The fivebrane giant graviton and its fluctuations}\label{ss:sphere}

We now study the spherical vacuum $\mathcal{S}$ in more detail. In the plane--wave, a spherical fivebrane in the six--plane can be stable, since the tension that tends to shrink it is balanced by its coupling to the six--form, as for the giant graviton of \cite{McGreevy:2000cw,Maldacena:2002rb}. In our model the tension is encoded in the five--bracket, while the coupling to the six--form is the term $\beta\det Z$, so the same balance should appear in the dynamics of the radius. To see it, we restrict to round \emph{dynamical} configurations of the six--plane,
\begin{eqn}\label{radialansatz}
 Z_a^m=R(t)\,\delta_a^{m-3},\qquad Y=H=\Theta=A=0 ,
\end{eqn}
which describe a five--sphere of radius $R(t)$. On this ansatz the potential and its derivative are
\begin{eqn}\label{app:radialfinite}
 V_R =\frac{R^2(R^4-\mu)^2}{12},\qquad
 V_R'=\frac{R(R^4-\mu)(5R^4-\mu)}6 .
\end{eqn}
Hence, the lagrangian and the equation of motion read as
\begin{eqn}\label{radial}
 L_R=3\dot R^2-\frac{R^2}{12}(R^4-\mu)^2,\qquad
 \ddot R+\frac{R(R^4-\mu)(5R^4-\mu)}{36}=0 .
\end{eqn}
The potential has two zero-energy radii, $R=0$ and $R_0=\mu^{1/4}$, separated by a stationary point at $R_s=(\mu/5)^{1/4}$. Hence, the stationary points associated to this potential are given by the radii
\begin{eqn}
    R=0\,,\qquad R_0^4 = \mu\,,\quad \text{and} \quad R_s^4 = \frac{\mu}{5}\,.
\end{eqn}
At $R_0$, both shifted brackets vanish, and the expanded configuration preserves all the dynamical supersymmetries. This is just the solution \eqref{spherevac}, and its worldvolume metric will be obtained from the parent M5 action in Section~\ref{sub:physicalM5}. Before studying the solution determined by $R_s$ we will now consider the fluctuation analysis around $\mathcal{S}$.

\smallskip

To study local stability and general fluctuations at $R_0$, we write $Z=R_0\mathbf1+\zeta$ and $Y=y$. Internal gauge rotations change the antisymmetric part of $\zeta$, which can therefore remove all its fifteen antisymmetric components. This leaves a symmetric matrix with a trace and twenty traceless components (while the linear Gauss constraint removes its conjugate momentum). 

We can now expand the potential around this solution obtaining
\begin{eqn}\label{spherequad}
 L_B^{(2)}={}&\frac12\tr\dot{\zeta}^2+\frac12\dot y^2+\frac12hF\dot h
 -\frac{\beta^2}{2}\|\tr(\zeta)\mathbf1-2\zeta\|^2
 -\frac{9\beta^2}{2}\|y\|^2-\frac{R_0^4}{4}h_{abc}h_{abc} .
\end{eqn}

\smallskip

For the tensor, combining this quadratic energy with its Dirac bracket gives
\begin{eqn}\label{app:tensoroscillator}
 \dot h=-\frac{R_0^4}{2}Fh,\quad \text{i.e.} \qquad 
 \dot h_{123}=-\frac\mu2h_{456},\qquad
 \dot h_{456}=\frac\mu2h_{123} .
\end{eqn}
Indeed, as discussed previously (see around eq. \eqref{dirac}) we should understand the tensor modes of different triples as complementary pair modes, as much as position and momentum of one harmonic oscillator. Each complementary pair is then one oscillator of frequency $\mu/2$, accounting for the ten tensor modes.

\smallskip

The fermion equation follows by evaluating $W_F$ on the sphere. In order to non separate the spinors along the directions $Y^i$ and $Z^m$ we define $\widehat\gamma_a=\gamma_{a+3}$ and write the fluctuation as $\Theta_a = \vartheta_a$. Clifford duality in six dimensions simply gives
\begin{eqn}\label{app:fermionlinear}
 (W_F)_{ab}=\beta G(\widehat\gamma_a\widehat\gamma_b-\delta_{ab}),\qquad
 \dot\vartheta_a=\beta G\left(\widehat\gamma_a\widehat\gamma_b\vartheta_b -\frac52\vartheta_a\right).
\end{eqn}
The operator $\widehat\gamma_a\widehat\gamma_b$ has eigenvalue six on this sixteen-dimensional subspace and zero on its eighty-dimensional complement. To be precise, it projects, up to normalization, onto the gamma trace space. Because $G^2=-1$, the EoM \eqref{app:fermionlinear} is a rotation with frequencies $7\beta/2$ and $5\beta/2$, respectively. Dividing each real dimension by two gives eight and forty fermion pairs, respectively in the gamma trace space and its complement. 

\medskip

Now, we can count the physical fluctuations associated to this solution
\begin{eqn}\label{spherespectrum}
 (n_B,\omega_B)&=(20,\mu/3),\ (1,2\mu/3),\ (18,\mu/2),\ (10,\mu/2),\\
 (n_F,\omega_F)&=(40,5\mu/12),\ (8,7\mu/12).
\end{eqn}
The bosonic entries refer, in order, to the traceless part of $\zeta$, its trace, the three--plane scalars $y$, and the chiral tensor. 
The two fermionic sectors are the gamma-traceless and gamma-trace parts of the six-vector spinor, with real dimensions eighty and sixteen. 

All physical bosonic frequencies are positive, establishing local stability of the sphere as expected. The weighted frequency sums also agree,
\begin{eqn}\label{zeropoint}
 \sum\omega_B=20\frac\mu3+28\frac\mu2+\frac{2\mu}3=\frac{64\mu}{3} 
 =40\frac{5\mu}{12}+8\frac{7\mu}{12}=\sum\omega_F.
\end{eqn}
so the bosonic and fermionic vacuum energies cancel in the quadratic expansion approximation.\footnote{Although there are different numbers of bosonic oscillators and fermion pairs, supersymmetry need not pair equal frequencies in a plane--wave: the time-dependent charges relate states with shifted energies, as also occurs in BMN \cite{Dasgupta:2002hx}.}


\smallskip

To end the fluctuation analysis, we can now check that $\mathcal S$ is invariant under $\mathfrak{su}(2|4)$, and its fluctuations organize into representations of this algebra. Indeed, in units of the lowest energy $\mu/12$, the modes in \eqref{spherespectrum} form a single multiplet $(SO(3), SO(6))_{\text{energy}}$,
\begin{eqn}\label{eq:fluctfivebrane}
(\mathbf1,\mathbf{20}')_4\to(\mathbf2,\mathbf{20})_5\to(\mathbf3,\mathbf6)_6\oplus(\mathbf1,\mathbf{10})_6\to(\mathbf2,\bar{\mathbf4})_7\to(\mathbf1,\mathbf1)_8 ,
\end{eqn}
where each step is the action of a supercharge (see eq. \eqref{Q}) $Q$ as $(\mathbf2,\mathbf4)$. This multiplet has a precise counterpart in BMN theory. Around the trivial vacuum $X=0$, the elementary oscillators of BMN theory, i.e.\ the six SO(6) scalars at $\frac\mu6$, the eight fermions at $\frac\mu4$ and the three SO(3) scalars at $\frac\mu3$, form the multiplet $(\mathbf1,\mathbf6)_2\to(\mathbf2,\mathbf4)_3\to(\mathbf3,\mathbf1)_4$ \cite{Dasgupta:2002hx}. In the language of \cite{Maldacena:2002rb}, this is the level $\ell\,=0$ of the fluctuations of the spherical transverse fivebrane. The higher levels are instead obtained from several oscillators and, in particular, the lowest state of level $\ell$ is the protected state built from $\ell+1$ SO(6) oscillators in the symmetric traceless representation, with energy $(l+1)\frac\mu6$. The matching of these states with the fluctuations of the fivebrane at large $N$ and strong coupling is indeed the main evidence that $X=0$ describes a fivebrane \cite{Maldacena:2002rb}. The multiplet \eqref{eq:fluctfivebrane} is exactly the level $\ell\,=1$ we described. Its lowest state, the $\mathbf{20}'$ at $\frac\mu3$, corresponds in BMN theory to a state of two SO(6) oscillators, while here it is an elementary fluctuation of the classical fivebrane. In other words, in BMN theory these modes are states containing several oscillators of the matrices, while in the $\mathcal{A}_6$ model they are elementary vibrations of the classical fivebrane. Since the lowest states of the multiplet are protected by supersymmetry, their energies are the same in the two descriptions. Together with the analysis below, all these ingredients aim to identify $\mathcal{S}$ as the fivebrane giant graviton of the plane--wave.

\paragraph{On the third stationary solution.} It is immediate to check that the intermediate stationary radius has positive energy and a negative radial Hessian. Using \eqref{radial} we obtain
\begin{eqn}\label{radialsaddle}
 V(R_s)=\frac{4\mu^{5/2}}{75\sqrt5},\qquad
 R=R_s+\rho:\quad \ddot\rho-\frac{4\mu^2}{45}\rho \equiv \ddot\rho- \tau_c^{-2} \rho  =0 .
\end{eqn}
It is therefore a static unstable configuration at the top of the radial barrier, whose perturbation has a growing component proportional to $\sim \exp[t/\tau_c]$, describing departure from this equilibrium.

In Euclidean time $\tau = it$, the same radial potential admits a finite-action trajectory between the collapsed and expanded zero-energy configurations.
Completing the Euclidean action into a square, gives
\begin{eqn}\label{app:euclideansquare}
 S_E&=\int d\tau\left[3(R')^2+\frac{R^2(R^4-\mu)^2}{12}\right]\\
 &=\int d\tau\,3\left(R'-\frac{R(\mu-R^4)}6\right)^2
   +\left[\frac\mu2R^2-\frac16R^6\right]_{\tau = -\infty}^{\tau = +\infty} .
\end{eqn}
Then,
\begin{eqn}\label{radialinstanton}
 R'=\frac{R(\mu-R^4)}6,\qquad
 R(\tau)^4=\frac{\mu}{1+e^{-2\mu(\tau-\tau_0)/3}}\,.
\end{eqn}
The trajectory approaches $R=0$ as $\tau\to-\infty$ and $R=R_0$ as $\tau\to+\infty$, with $\tau_0$ specifying its center.
Finally, evaluating the boundary term at $R=0$ and $R=\mu^{1/4}$ gives
\begin{eqn}\label{app:instantonaction}
 S_E=\frac12\mu\,\mu^{1/2}-\frac16\mu^{3/2}
     =\frac{\mu^{3/2}}3 \,.
\end{eqn}
\begin{figure}[htbp]
 \centering
 \includegraphics[width=0.86\linewidth]{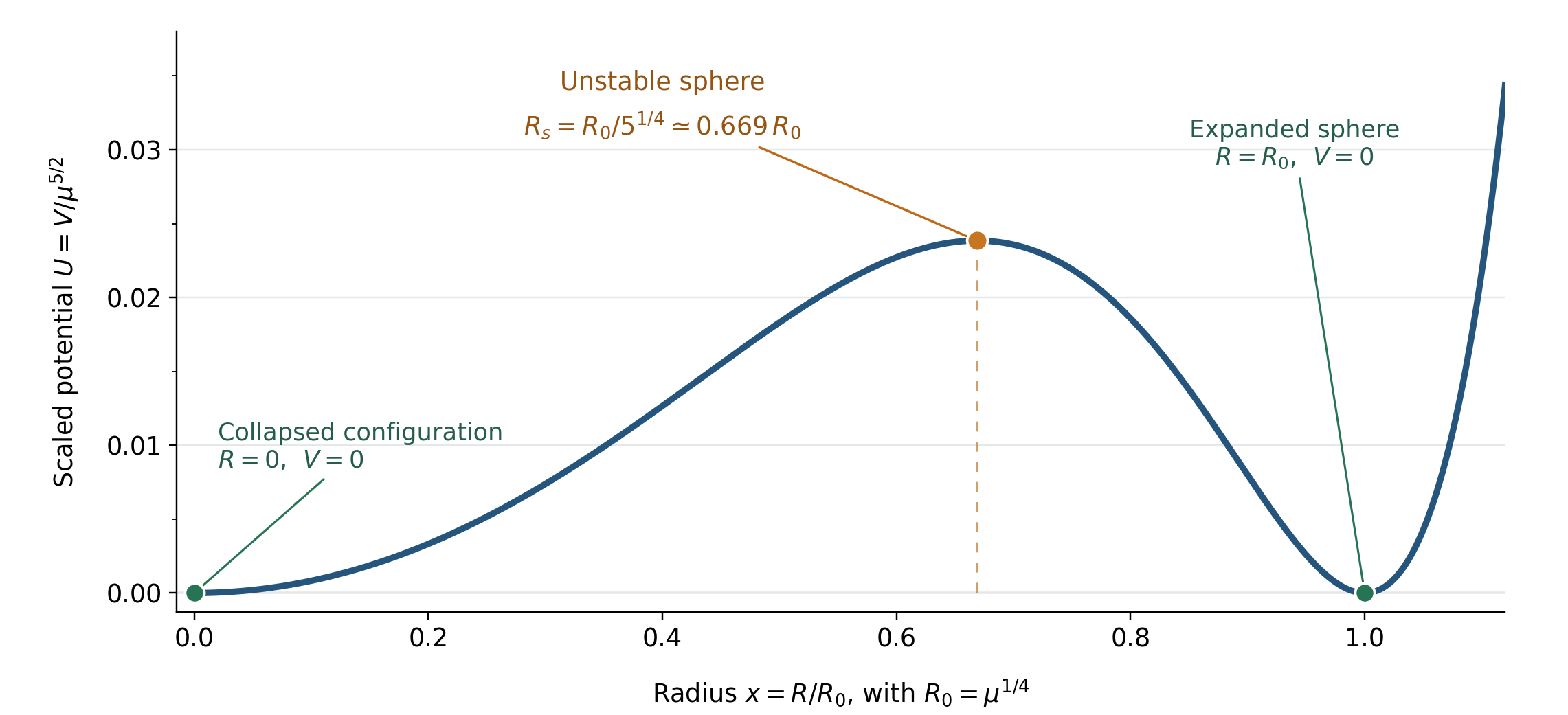}
 \caption{The radial potential of equation \eqref{radial}. The collapsed and expanded zero-energy configurations are separated by the unstable stationary sphere at $R_s=(\mu/5)^{1/4}$.}
 \label{fig:radialpotential}
\end{figure}

\subsection{The $\mathcal{R}$ vacua and their fluctuations}
\label{ys:section}
For the spherical vacuum, the $Y^i$ scalars and the tensor vanish, and the dynamics is entirely controlled by the six--plane scalars, whose tension is balanced by the six--form coupling. We now consider the opposite situation, where the six--plane scalars vanish, $Z=0$, and the relevant fields are $Y^i$. In this case the five--bracket vanishes identically, and the potential reduces to the square of the shifted two--bracket. The important difference is that this bracket is built with the dynamical tensor $H$. Hence a change of the radius of the scalars also changes the tensor, and the scalar equation alone is not enough to determine an equilibrium.

\medskip

To start, let us recall the bosonic action of \eqref{Lmass} at $Z=0$
\begin{equation}\label{ys:Lgeneral}
 L_{Y}=\frac12(D_tY_a^i)^2+\frac12H_{abc}F_{abcdef}D_tH_{def}
 -\frac14\bigl(H_{abc}Y_b^iY_c^j-\alpha\,\epsilon^{ijk}Y_a^k\bigr)^2 .
\end{equation}
For simplicity we restrict to round \emph{dynamical} configurations of the three--plane,
\begin{equation}\label{ys:ansatzY}
 Y_i^{\ j}=r(t)\,\delta_i^{\ j},\qquad Y_A^{\ j}=0,\qquad Z=\Theta=0,
\end{equation}
where $A=4,5,6$ labels again the unoccupied triple. Since the tensor must now evolve together with the radius, we also need an ansatz for $H$. Motivated by the two families found in Section~\ref{ss:ranks}, we take
\begin{equation}\label{ys:branches}
 H_{123}=h(t),\qquad H_{456}=k(t),\qquad H_{ijA}=0,\qquad
 H_{iAB}=\chi\,h\,\epsilon_{AB,i+3},
\end{equation}
with $\chi=0$ for $\mathcal{R}_{\rm I}$ and $\chi=1$ for $\mathcal{R}_{\rm II}$.
Within this ansatz and in the temporal gauge, the only surviving two--brackets are $B_i^{jk}$, and the tensor kinetic term reduces to $3(h\dot k-k\dot h)$, since $H_{123}$ and $H_{456}$ form a canonical pair. Up to a total derivative, the reduced action and Hamiltonian are
\begin{equation}\label{ys:Lradial}
 L_3=\frac32\dot r^2+6h\dot k-V_3(r,h),\qquad \mathcal H_3=\frac{p_r^2}{6}+V_3(r,h),
\end{equation}
with
\begin{equation}\label{ys:Hradial}
 p_r=3\dot r,\qquad V_3(r,h)=\frac32r^2(hr-\alpha)^2\,.
\end{equation}
Notice that $\chi$ has disappeared, so the two branches have the same radial dynamics, although they are embedded differently in the full theory. Varying \eqref{ys:Lradial} with respect to $r$, $k$ and $h$ gives\footnote{More generally, one can keep independent diagonal deformations $r_i$. With the same tensor ansatz the Lagrangian becomes
\begin{equation}\label{ys:Ldiagonal}
 L_{\mathrm{diag}}=\frac12\sum_{i=1}^3\dot r_i^2+6h\dot k-\frac12\sum_{\mathrm{cyc}}(hr_jr_k-\alpha r_i)^2\,,
\end{equation}
where $(i,j,k)$ is a cyclic permutation. It admits for instance the solution $r_1=A\cos(\alpha t+\varphi)$, $r_2=r_3=0$, i.e.\ an exact oscillation along a single direction, with constant $h$ and $k$.}
\begin{equation}\label{ys:radialeom}
 \ddot r=-r(hr-\alpha)(2hr-\alpha),\qquad \dot h=0,\qquad \dot k=\frac12r^3(hr-\alpha).
\end{equation}
Hence the flux $h$ is conserved, as the momentum conjugate to $k$, which does not appear in the potential. The radius therefore moves in the potential $V_3$ at fixed $h$, while $k$ evolves in time, unless the configuration sits at $hr=\alpha$.

\medskip

We can now look for the stationary points, as we did for the spherical vacuum in Section~\ref{ss:sphere}. At fixed $h$ the radius moves in the one--dimensional potential $V_3$, whose stationary points are
\begin{equation}\label{ys:radialcritical}
 r=0,\qquad r_0=\frac\alpha h,\qquad r_s=\frac\alpha{2h}\,.
\end{equation}
At the first two radii the potential vanishes, and by \eqref{ys:radialeom} also $\dot k=0$, so these are static supersymmetric solutions, namely the collapsed configuration and the vacuum $\mathcal{R}$ with $hr_0=\alpha$. The intermediate radius, instead, is the top of the barrier between them, with positive energy,
\begin{eqn}
 V_3(r_s)=\frac{3\alpha^4}{32h^2} \,,
\end{eqn}
and it is unstable.\footnote{For $r=r_s+\rho$ the radial equation gives $\ddot\rho=\frac{\alpha^2}{2}\rho$, so that the perturbation grows as $e^{t/\tau_c}$ with $\tau_c^{-1}=\frac\alpha{\sqrt2}=\frac\mu{3\sqrt2}$.} Differently from the spherical case, however, this solution is not static. Since $hr_s\neq\alpha$, the last equation in \eqref{ys:radialeom} makes the flux $k$ evolve linearly in time,
\begin{equation}\label{ys:driftingsaddle}
 r(t)=r_s,\qquad h(t)=h,\qquad k(t)=k_0-\frac{\alpha^4}{32h^3}\,t ,
\end{equation}
while the radius stays at the top of the barrier. In other words, this configuration is held in place by a flux that changes in time.\footnote{\label{foot:klinear}One could think that a linear growth of $k$ is a pathology. Since $k$ does not appear in the Hamiltonian, the energy of the configuration remains finite, and the drift only reflects the fact that $k$ is the variable conjugate to the conserved flux $h$. In particular, in a state with definite $h$ the variable $k$ is completely undetermined, and its drift becomes a phase.}

\smallskip

As for the spherical vacuum, a Euclidean trajectory connecting the collapsed and the expanded configurations can be obtained by completing a square. Here, however, we must first fix the flux $h$, which is conserved. Subtracting $P_k\dot k$ from \eqref{ys:Lradial} and continuing to Euclidean time, $t=-\ii\tau$, we obtain
\begin{equation}\label{ys:Euclidean}
\begin{aligned}
S_{E,h}&=\int d\tau\left[\frac32(r')^2+\frac32r^2(\alpha-hr)^2\right]\\
&=\frac32\int d\tau\left[r'-r(\alpha-hr)\right]^2+\left[\frac{3\alpha}{2}r^2-hr^3\right]_{-\infty}^{+\infty}.
\end{aligned}
\end{equation}
The action is therefore minimized by the first--order equation
\begin{equation}\label{ys:instanton}
r'=r(\alpha-hr),\qquad r(\tau)=\frac{r_0}{1+e^{-\alpha(\tau-\tau_0)}}\,,
\end{equation}
whose solution goes from the collapsed configuration at $\tau\to-\infty$ to the expanded one at $\tau\to+\infty$. Its action is given by the boundary term,
\begin{equation}\label{ys:instantonaction}
S_{E,h}^{\mathrm{inst}}=\frac{\alpha^3}{2h^2}=\frac{\mu^3}{54h^2}.
\end{equation}

These formulae are essentially the same as for the spherical vacuum, but there is a new feature when we lift the trajectory to the tensor phase space. Indeed, the Euclidean kinetic term of the tensor is $-6\ii hk'$, so that the $h$ equation gives $k'=\frac\ii2r^3(\alpha-hr)$. Along \eqref{ys:instanton} this integrates to
\begin{equation}\label{ys:complexlift}
k(\tau)=k_-+\frac{\ii}{6}r(\tau)^3,
\end{equation}
and the lifted trajectory is complex in the variable $k$. As explained already in Footnote \ref{foot:klinear}, since we are considering conserved $h$, \eqref{ys:Euclidean} defines the problem at fixed $h$, and the corresponding amplitude is computed in the $h$--representation, where $k$ is never fixed at the endpoints. The complex value of $k$ along the trajectory has therefore no physical meaning. In other words, \eqref{ys:instantonaction} is the action at fixed flux $h$, and not the action of a transition between two given real values of $k$.
Ultimately, both Euclidean trajectories are the analogue of the instantons connecting different vacua of the BMN model \cite{Yee:2003ge}, and of the tunnelling between point--like and giant gravitons \cite{Grisaru:2000zn}. As in those cases, whether they carry fermion zero modes, and therefore whether they lift the classical degeneracy of the zero--energy configurations, requires a separate analysis, which we leave for future work.

\paragraph{Fluctuations around the supersymmetric solution.}
We now expand around either branch of the supersymmetric solution with radius $R=r_0$. At this point all covariant velocities are zero, and consequently the fermion variation therefore vanishes for all sixteen dynamical supersymmetries, as for the $\mathcal{S}$ solution. 
We then write the most general scalar and tensor fluctuations according to the ansatz \eqref{ys:ansatzY}--\eqref{ys:branches} as
\begin{equation}\label{ys:fluctansatz}
\begin{aligned}
Y&=\begin{pmatrix}r_0 \mathbf{1}_3+\sigma\\0\end{pmatrix},\qquad Z=\begin{pmatrix}U\\V\end{pmatrix},\\
H_{123}&=h+\eta,\qquad H_{456}=k_0+\kappa,\qquad H_{ijA}=\epsilon_{ijk}t_{k,A-3},\\
H_{iAB}&=\epsilon_{ABC}\bigl(\chi h\delta_{i,C-3}+m_{i,C-3}\bigr).
\end{aligned}
\end{equation}
First, as for the solution $\mathcal{S}$, the gauge redundancy removes the three antisymmetric upper components of $Y$ (and its nine lower components), so that only the six entries of $\sigma$ remain.
The simplest evaluation of the fluctuation around the solution is due to the scalar potential. One directly obtains
\begin{equation}\label{ys:VYquad}
V_Y^{(2)}=\frac12\left\|\alpha(\operatorname{tr}\sigma\,\mathbf{1}_3-2\sigma)+r_0^2\eta\mathbf{1}_3\right\|^2.
\end{equation}
Thus, at fixed $\eta$, the trace gives one oscillator of frequency $\alpha$, while the five traceless components have frequency $2\alpha$. 
For the six-plane scalars, the quadratic potential is
\begin{equation}\label{ys:Zquad}
V_Z^{(2)}=\frac12(2\alpha^2+\beta^2)\|U\|^2+\frac12(2\chi^2\alpha^2+\beta^2)\|V\|^2.
\end{equation}
There are eighteen entries in each matrix. The upper entries always have frequency $\mu/2$ for each branch. The lower entries have frequency $\mu/6$ on branch I and $\mu/2$ on branch II. 
Finally, for the tensor we must first solve the mixed Gauss constraint \eqref{gauss}. It is then useful to keep the full mixed matrix $M$ and write $H_{ijA}$ explicitly. Directly contracting the six-dimensional epsilon gives
\begin{equation}\label{ys:Kexact}
\begin{aligned}
K_{iA}&=2\bigl(\operatorname{cof}M+\operatorname{cof}t-hM-kt\bigr)_{i,A-3},\\
K_{ij}&=2(Mt^T-tM^T)_{ij},\qquad K_{AB}=2(M^Tt-t^TM)_{A-3,B-3}.
\end{aligned}
\end{equation}
We can indeed check that it returns \eqref{eq:K1} for $t =0$.
Expanding this expression around \eqref{ys:fluctansatz}, the mixed Gauss equation reads (see eq. \eqref{gauss})
\begin{equation}\label{ys:Gausslinear}
P_{Ai}=\frac3R\delta K_{iA},\qquad \delta K=2\left\{h\bigl[\chi\operatorname{tr}m\,\mathbf{1}_3-\chi m^T-m-\chi\eta\mathbf{1}_3\bigr]-k_0t\right\}.
\end{equation}
Combining it with the lower two-bracket potential gives
\begin{equation}\label{ys:Htensorquad}
\mathcal H_{m,t}^{(2)}=\frac{r_0^4}{2}\|t\|^2+\frac{18}{r_0^2}\left\|h\bigl[\chi (m^T+\eta\mathbf{1}_3-\operatorname{tr}m\,\mathbf{1}_3)+m\bigr]+k_0t\right\|^2.
\end{equation}

Hence, there are nine tensor oscillators at $\mu/3$ on branch I. On branch II there is one at $\mu/3$ and five at $2\mu/3$.
The remaining variable $\kappa$ is absent from the quadratic Hamiltonian, so by \eqref{dirac} $\eta$ is constant at this order.

The fermion equation is obtained by evaluating the Yukawa operator on the same background. Since $W_F=0$, one has
\begin{equation}\label{ys:fermioneom}
\dot\Theta=-\left(W_H+\frac\mu4G\right)\Theta,\qquad (W_H)_{ij}=\alpha\epsilon_{ijk}\gamma_k,\qquad (W_H)_{AB}=\chi\alpha\epsilon_{ABC}\gamma_{C-3}.
\end{equation}
There is no mixing between the two internal triples, and $k_0$ does not enter because the complementary scalar background vanishes. For the occupied triple, the Clifford identity $\epsilon_{ijk}\gamma_k=-(\gamma_i\gamma_j-\delta_{ij})G$ gives
\begin{equation}\label{ys:fermionprojector}
\dot\theta_i=G\left(\alpha\gamma_i\gamma_j\theta_j-\frac{7\mu}{12}\theta_i\right),\qquad (P_{\mathrm{tr}})_{ij}=\frac13\gamma_i\gamma_j.
\end{equation}
The gamma-trace projector has real rank sixteen, and its complement has real rank thirty-two. Since $G^2=-1$, their frequencies are $5\mu/12$ and $7\mu/12$, respectively. Dividing each real dimension by two gives eight and sixteen fermion pairs. 
On branch I the complementary Yukawa block vanishes, leaving twenty-four fermion pairs of frequency $\mu/4$. On branch II it repeats the occupied block. We can now collect the nonzero frequencies as
\begin{equation}\label{ys:spectraI}
\begin{aligned}
\mathcal R_{\mathrm I}:\qquad (n_B,\omega_B)&=(10,\mu/3),\ (5,2\mu/3),\ (18,\mu/2),\ (18,\mu/6),\\
(n_F,\omega_F)&=(8,5\mu/12),\ (16,7\mu/12),\ (24,\mu/4),
\end{aligned}
\end{equation}

\begin{equation}\label{ys:spectraII}
\begin{aligned}
\mathcal R_{\mathrm{II}}:\qquad (n_B,\omega_B)&=(2,\mu/3),\ (10,2\mu/3),\ (36,\mu/2),\\
(n_F,\omega_F)&=(16,5\mu/12),\ (32,7\mu/12).
\end{aligned}
\end{equation}
Both lists must be supplemented by the collective zero pair along the vacuum family, i.e. the canonical pair $(h,k)$. 
Finally, all displayed bosonic frequencies are positive, while the weighted sums satisfy
\begin{equation}\label{ys:frequencysums}
\sum\omega_B^{\mathrm I}=\sum\omega_F^{\mathrm I}=\frac{56\mu}{3},\qquad \sum\omega_B^{\mathrm{II}}=\sum\omega_F^{\mathrm{II}}=\frac{76\mu}{3}.
\end{equation}
Hence the bosonic and fermionic zero-point contributions cancel in the quadratic approximation, in agreement with the six-dimensional spherical solution in \eqref{zeropoint}. 

The $\mathcal{R}$ vacua are also $\mathfrak{su}(2|4)$ invariant. The $SO(6)$ rotations act trivially since $Z=0$, while an $SO(3)$ rotation is compensated by a gauge rotation of the first triple on $\mathcal R_{\rm I}$, and of both triples on $\mathcal R_{\rm II}$, which leaves the tensor invariant. Correspondingly, the modes of the occupied triple \eqref{ys:spectraII} form the multiplet
\begin{eqn}
(\mathbf1,\mathbf1)_4\to(\mathbf2,\mathbf4)_5\to(\mathbf3,\mathbf6)_6\to(\mathbf4,\bar{\mathbf4})_7\to(\mathbf5,\mathbf1)_8 ,
\end{eqn}
which is the multiplet of the $\mathfrak{su}(2)$ BMN model around its fuzzy sphere \cite{Dasgupta:2002hx}. On $\mathcal R_{\rm I}$ the second triple gives three copies of the multiplet $(\mathbf1,\mathbf6)_2\to(\mathbf2,\mathbf4)_3\to(\mathbf3,\mathbf1)_4$ of the trivial BMN vacuum, where the tensor modes replace the three--plane scalars. On $\mathcal R_{\rm II}$, instead, it gives a second copy of the fuzzy--sphere multiplet, in agreement with the diagonal embedding of $\mathfrak{su}(2)\oplus\mathfrak{su}(2)$ found in Section~\ref{ss:tensoralgebra}. In BMN language, $\mathcal R_{\rm I}$ therefore resembles a fuzzy sphere together with three trivial blocks, while $\mathcal R_{\rm II}$ resembles two fuzzy spheres.

\section{The plane--wave M5 and its spherical sector}\label{sec:m5}

The massive $\mathcal{A}_6$ model was constructed in Section~\ref{sec:mass} by requiring supersymmetry, but this construction does not fully explain the physical origin of each term. In this section we give it a geometric origin, by deriving it from the light--cone M5--brane in the plane--wave at its lowest harmonics, including the self--dual tensor and the fermions. We proceed in three steps. First, as exercise we review the bosonic spherical fivebrane of \cite{Maldacena:2002rb}, which fixes the physical radius and shows how the flux stabilizes the sphere, in analogy with the $\mathcal{S}$ vacuum found in Section \ref{sec:vacua}. Then we write the light--cone Hamiltonian of the M5--brane in the plane--wave, reviewing the formulation of \cite{Bergshoeff:1998vx,Bandos:2008fr}, which keeps the self--dual tensor and makes supersymmetry manifest. Finally, we perform the finite spherical reduction of \cite{Artime:2026mph}. We will see that the background produces exactly the masses and the polarization terms fixed by supersymmetry in Section~\ref{sec:mass}. The detailed superspace and canonical computations are collected in Appendix~\ref{app:superM5}.

\subsection{The spherical M5 in the plane--wave}\label{sub:physicalM5}

We start by reviewing the bosonic spherical fivebrane of \cite{Maldacena:2002rb}. The goal is twofold. On the one hand, it relates the radius of the sphere to the light--cone momentum and to the M5 tension, which are absorbed in the normalization of the finite model. On the other hand, it shows that already for the bosonic brane the energy of the sphere is a perfect square, which is the simplest case of \eqref{Vsquares} in absence of dynamical self--dual tensor and only considering the six--dimensional plane.

In our conventions for the flux and the orientation, the maximally supersymmetric plane--wave is \cite{Berenstein:2002jq,Dasgupta:2002hx,Maldacena:2002rb}
\begin{eqn}\label{ppwave}
 ds^2&=-2dx^+dx^-+dX^IdX^I-U(X)(dx^+)^2,\qquad
 U=\Bigl(\frac\mu3\Bigr)^2Y^2+\Bigl(\frac\mu6\Bigr)^2Z^2,\\
 F_{4,B}&=-\mu\,dx^+\wedge\vol_3,\qquad
 F_{7,B}=*_{11}F_{4,B}=+\mu\,dx^+\wedge\vol_6 .
\end{eqn}
The function $U$ gives different masses to the three--plane and to the six--plane, while the four--form flux and its dual seven--form are the fields to which membranes and fivebranes couple. With the tensor and the fermions set to zero, the covariant M5 action \cite{Pasti:1997gx,Bandos:1997ui,Aganagic:1997zq} reduces to the volume term and to the coupling to the six--form potential,
\begin{eqn}\label{app:bosonicM5}
 S_B=-T_5\int d^6\sigma\sqrt{-\det g_{\alpha\beta}}
       +T_5\int C_{6,B} ,\qquad T_5=\frac1{(2\pi)^5\ell_p^6}.
\end{eqn}
We now consider a round fivebrane in the six--plane, $Y=0$ and $Z^m=R(t)\,u_{m-3}$, in the gauge $x^+=t$. The induced metric is
\begin{eqn}\label{app:radialmetric}
 g_{ij}=R^2\gamma_{ij},\qquad g_{0i}=0,\qquad
 -g_{00}=D \equiv 2\dot x^-+\frac{\mu^2R^2}{36}-\dot R^2 ,
\end{eqn}
where $\gamma$ is the metric of the unit five--sphere. Inserting it in \eqref{app:bosonicM5}, with the potential $c_{5,B}$ of \eqref{electricprimitives}, we obtain
\begin{eqn}\label{app:radialunfixed}
 L_B=-T_5\Omega_5R^5\sqrt D+\frac{\mu T_5\Omega_5}{6}R^6,
 \qquad \pi_-\equiv\frac{\partial L_B}{\partial\dot x^-}
       =-\frac{T_5\Omega_5R^5}{\sqrt D}=-p\,\Omega_5 .
\end{eqn}
The last equality fixes the light--cone momentum, $P^+_{\rm tot}=p\,\Omega_5$, and it determines the longitudinal velocity in terms of the radial motion,
\begin{eqn}\label{app:longitudinal}
 D=\frac{T_5^2R^{10}}{p^2},\qquad
 \dot x^-=\frac12\left(\frac{T_5^2R^{10}}{p^2}
                   -\frac{\mu^2R^2}{36}+\dot R^2\right).
\end{eqn}
Since the light--cone theory describes a sector of fixed momentum, we can remove $x^-$ by a partial Legendre transform. This gives the Routhian \cite{Maldacena:2002rb}
\begin{eqn}\label{app:radialexpanded}
 L^{M5}_R = L_B-\pi_-\dot x^-
 &=\frac{p\Omega_5}{2}\dot R^2
 -\frac{\Omega_5T_5^2}{2p}R^{10}
 -\frac{p\Omega_5\mu^2}{72}R^2
 +\frac{\mu T_5\Omega_5}{6}R^6  \\
 &= \frac{p\Omega_5}{2}\dot R^2
 -\frac{\Omega_5}{2p}\left(T_5R^5-\frac{\mu p}{6}R\right)^2 .
\end{eqn}
In this perspective the three terms of the potential have a clearer explanation than \eqref{app:radialfinite}. The $R^{10}$ term is the energy of the tension, which tends to shrink the sphere, the $R^2$ term comes from the mass of the six--plane in $U$, and the negative $R^6$ term is the coupling to the six--form, which tends to expand it. The three combine into a perfect square, as indeed predicted by the spherical vacuum in Section~\ref{ss:sphere}. Hence, besides the collapsed configuration, the potential vanishes at
\begin{eqn}\label{radiusphysical}
 R_0^4=\frac{\mu p}{6T_5}
       =\frac{\mu P^+_{\mathrm{tot}}}{6T_5\Omega_5} ,
\end{eqn}
where the tension is exactly balanced by the flux. At this radius the two terms in \eqref{app:longitudinal} cancel, so $\dot x^-=0$, and the induced metric is
\begin{eqn}\label{induced}
 ds^2_{\mathrm{wv}}=R_0^2\left[-\left(\frac\mu6\right)^2dt^2+d\Omega_5^2\right] .
\end{eqn}
Finally, it is worth to mention that \eqref{app:radialfinite} and \eqref{radiusphysical} coincide once $p$ and $T_5$ are absorbed in the normalization of the finite model, where $R_0^4=\mu$. Hence $\mathcal{S}$ is the finite counterpart of the spherical fivebrane of \cite{Maldacena:2002rb}, and \eqref{radiusphysical} restores the dependence of its radius on the light--cone momentum.

\subsection{The light--cone worldvolume action}\label{sub:LCM5}

The bosonic action \eqref{app:bosonicM5} is enough to describe the shape of the fivebrane, but it does not contain the fermions. To obtain their couplings to the background, we have to use the supersymmetric version of the M5 action, where the background enters through superfields. In particular, the metric is replaced by the supervielbein $E^a$, and the potentials by the superforms $C_3$ and $C_6$ \cite{Bandos:1997ui,Aganagic:1997zq,Claus:1998fh}. These superfields depend both on the bosonic coordinates and on a thirty--two component Grassmann coordinate $\vartheta$, and their expansion in $\vartheta$ determines how the worldvolume fermions couple to the plane--wave.

However, not all the components of $\vartheta$ are physical. The action has a local fermionic symmetry, kappa symmetry, which removes half of them. As for the membrane, the most convenient choice is the light--cone gauge,
\begin{eqn}\label{kappagauge}
 \G^+\vartheta=0 ,\qquad \vartheta=2^{-1/4}(0,\psi)^T ,
\end{eqn}
which leaves a single sixteen--component real spinor $\psi$. The great advantage of this gauge is that the expansion in the fermions becomes very short. Indeed, in the plane--wave the vector supervielbein stops at quadratic order in the fermions, as shown in \cite{Dasgupta:2002hx} using the superspace construction of \cite{Claus:1998fh}, and the spinor vielbein is linear in $\psi$. In our conventions we find the exact expressions
\begin{eqn}\label{exactvielbein}
 E^+=dt,\qquad E^I=dX^I,\qquad
 E^-=dx^-+\tfrac12U\,dt-\ii\psi^Td\psi-\frac{\ii\mu}{4}\psi^TG\psi\,dt .
\end{eqn}
Hence the only fermionic corrections appear in $E^-$, i.e.\ in the longitudinal direction. This is important, since we saw in Section~\ref{sub:physicalM5} that, once the light--cone momentum is fixed, the coefficient of $dx^-$ is replaced by $p$. In the same way, the two fermionic terms in $E^-$ become the kinetic term and the mass term of the worldvolume fermions, multiplied by $p$. In particular, the mass $\frac\mu4$ comes from the spinor connection of the plane--wave, acting on the sixteen physical components selected by \eqref{kappagauge}.

As explained in Appendix \ref{app:superM5} the same connection acts also on the other half of the thirty--two components, which describes the supersymmetry parameters. There it gives a rotation with frequency $\frac\mu{12}$, which is precisely the time dependence $\epsilon(t)=e^{\mu tG/12}\epsilon_0$ found in Section~\ref{sec:mass}. In other words, the time dependence that we obtained from the Noether procedure is simply the time dependence of the Killing spinors of the plane--wave.\footnote{The same analysis applies to the remaining half of the Killing spinors, which generate the kinematical supersymmetries. In this case the Killing spinor equation gives $\eta(t)=e^{-\mu tG/4}\eta_0$, which reproduces the sixteen kinematical supersymmetries of the centre of mass introduced in \eqref{eq:kinsusy}.}

\smallskip

Besides the supervielbein, the fivebrane couples to the background through the potentials $C_3$ and $C_6$, which in the bosonic case gave the polarization term in \eqref{app:radialunfixed}. Their fermionic completion can be obtained by integrating the corresponding superform field strengths along the fermionic coordinates, following \cite{Claus:1998fh}. In the light--cone gauge \eqref{kappagauge} the result is again very simple,
\begin{eqn}\label{exactpotentials}
 C_3&=dt\wedge\left(c_{2,B}-\ii\,dX^I\wedge\psi^T\gamma_Id\psi\right),\\
 C_6&=dt\wedge\left(c_{5,B}+\frac\ii{4!}dX^I\wedge dX^J\wedge dX^K\wedge dX^L
                                   \wedge\psi^T\gamma_{IJKL}d\psi\right) ,
\end{eqn}
where the bosonic parts are
\begin{eqn}\label{electricprimitives}
 c_{2,B}=\frac\mu3\,\iota_Y\vol_3,\qquad
 c_{5,B}=-\frac\mu6\,\iota_Z\vol_6 ,
\end{eqn}
and $\iota_Y$, $\iota_Z$ denote the contraction with the radial vector of the three--plane and of the six--plane. The bosonic parts reproduce the fluxes of the plane--wave. Indeed, since $d(\iota_Y\vol_3)=3\vol_3$ and $d(\iota_Z\vol_6)=6\vol_6$, we find $d(dt\wedge c_{2,B})=-\mu\,dt\wedge\vol_3=F_{4,B}$ and $d(dt\wedge c_{5,B})=\mu\,dt\wedge\vol_6=F_{7,B}$, as in \eqref{ppwave}. The fermionic parts, instead, are new. Once pulled back to the worldvolume, the term in $C_3$ couples the fermions to one scalar and, through the tensor, gives the Yukawa coupling $W_H$, while the term in $C_6$ couples them to four scalars and gives the five--bracket Yukawa coupling $W_F$.\footnote{Notice that both fermionic terms are proportional to $dt$, so that no higher powers of the fermions can appear, and \eqref{exactpotentials} is exact.}

\medskip
\medskip

We can now write the light--cone action of the fivebrane in the plane--wave. We insert the superfields \eqref{exactvielbein} and \eqref{exactpotentials} into the Hamiltonian M5 action of \cite{Bergshoeff:1998vx}, and we impose the constraint which expresses the chirality of the tensor. In flat space, the same procedure leads to the light--cone action of \cite{Bandos:2008fr}, and here we only have to include the background.

To do this, we split the worldvolume $\mathcal{W}$ into time and a five--dimensional spatial slice $\Sigma$, and we fix the light--cone momentum density to $p\,w$, where $w\,d^5\sigma$ is the measure of the unit sphere. Since $C_3$ in \eqref{exactpotentials} is proportional to $dt$, its pullback to $\Sigma$ vanishes, and the spatial part of the field strength is simply $h=d_\Sigma A_2$. It is then convenient to introduce the induced spatial metric, the bivector dual to $h$, and the bracket that it defines,
\begin{eqn}\label{tensorbivector}
 g_{ij}=\partial_iX^I\partial_jX^I,\qquad
 b^{ij}=\frac{\epsilon^{ijklm}h_{klm}}{3!\,w},\qquad
 \Bh(f,g)=b^{ij}\partial_if\,\partial_jg .
\end{eqn}
The bracket $\Bh$ contracts two gradients with the flux, while the Nambu bracket $\N$, used below \eqref{fullM5},
\begin{eqn}
    \N(f_1, f_2,f_3,f_4,f_5) = \frac{\epsilon^{ijklm}}{w}  \partial_if_1\,\partial_j f_2\, \partial_kf_3\,\partial_l f_4 \, \partial_m f_5\,.
\end{eqn}
is the Jacobian of five functions. We will see that these are the worldvolume counterparts of the two--bracket and of the five--bracket of the matrix model, and they allow us to write the tensor energy and the volume energy in the same way.

Eliminating the transverse momenta, as shown in Appendix~\ref{app:canonicalM5}, we obtain the worldvolume Lagrangian\footnote{\label{vvector}The residual gauge field is a vector $v^i$ which preserves the volume, $\partial_i(wv^i)=0$, and enters through $D_tf=\partial_tf-v^i\partial_if$ and $D_tA_2=\partial_tA_2-\iota_vh$ \cite{Bandos:2008fr}.}
\begin{eqn}\label{fullM5}
 L_{\mathrm{M5}}={}&\int_\Sigma pw\left[
 \frac 12(D_tX)^2-\frac 12U
 -\frac{T_5^2}{2p^2\,5!}\N(X^{I_1},\ldots,X^{I_5})^2
 -\frac{T_5^2}{4p^2}g_{ik}g_{jl}b^{ij}b^{kl}\right]\\
 &+\frac{T_5}{2}\int_\Sigma D_tA_2\wedge h
 +T_5\int_\Sigma(c_{5,B}-h\wedge c_{2,B})\\
 &+\ii p\int_\Sigma w\,\psi^TD_t\psi
 +\ii T_5\int_\Sigma w\,\psi^T\gamma_I\Bh(X^I,\psi)\\
 &+\frac{\ii T_5}{4!}\int_\Sigma w\,\psi^T\gamma_{IJKL}
                       \N(X^I,X^J,X^K,X^L,\psi)
 +\frac{\ii\mu p}{4}\int_\Sigma w\,\psi^TG\psi ,
\end{eqn}
where integrals containing $w$ include $d^5\sigma$. Each line has a simple meaning. The first line contains the kinetic energy of the embedding, the potential $U$ of the plane--wave, the energy of the volume and the energy of the tensor flux. The second line contains the first--order kinetic term of the tensor, which describes its chiral phase space, and the couplings to the background. Here the six--form coupling $c_{5,B}$ is the one we met for the bosonic sphere, while the coupling $h\wedge c_{2,B}$ of the tensor to the three--form arises when the chirality constraint combines the momentum of the tensor with the Wess--Zumino term. The last two lines contain the fermions. The kinetic and mass terms come from $E^-$, while the two Yukawa couplings come from the fermionic parts of $C_3$ and $C_6$. Finally, for $\mu=0$ we recover the flat light--cone action of \cite{Bandos:2008fr}, and its spherical reduction in \cite{Artime:2026mph}.

\smallskip

Before concluding this part, let us comment on the gauge symmetry of \eqref{fullM5}. The light--cone gauge does not fix the reparametrizations of $\Sigma$ which preserve the density $w$, i.e.\ the volume--preserving diffeomorphisms $\mathrm{SDiff}(\Sigma)$, whose gauge field is the vector $v^i$ of footnote~\ref{vvector}. Notice that the mass and polarization terms do not contain covariant derivatives, so they do not depend on $v^i$. Hence the constraint obtained by varying $v^i$ is the same as in the massless theory \cite{Bergshoeff:1998vx,Bandos:2008fr}. In our notation it reads
\begin{eqn}\label{M5Gauss}
 d_\Sigma\mathscr K=0,\qquad
 \mathscr K=\Pi_I\,dX^I+\frac{T_5}{4}h_{ijk}b^{jk}\,d\sigma^i +\ii p\,\psi^Td\psi ,
\end{eqn}
where $\Pi_I$ is the transverse momentum density with respect to the measure $w\,d^5\sigma$. The one--form $\mathscr K$ is the momentum density along $\Sigma$, with the contributions of the scalars, of the tensor and of the fermions, and the constraint states that it is closed, i.e.\ that it has no component which can generate a volume--preserving diffeomorphism. On the spherical ansatz \eqref{M5dictionary}, $\mathscr K$ takes the form
\begin{eqn}\label{M5Gausssphere}
 \frac{\mathscr K}{p} = -\Ga_{ab}\,u_a\,du_b + d_\Sigma\chi ,
\end{eqn}
where $\chi$ is a function which drops out of the constraint. Hence $d_\Sigma\mathscr K=-p\,\Ga_{ab}\,du_a\wedge du_b$, and since the two--forms $du_a\wedge du_b$ are independent on the sphere, the worldvolume constraint is equivalent to the Gauss' law 
\begin{eqn}
    \Ga_{ab}=0 \iff d_\Sigma \mathscr{K} = 0
\end{eqn}
of the finite model. In other words, the volume--preserving diffeomorphisms of the fivebrane reduce, on the lowest harmonics, to the $SO(6)$ gauge symmetry of the $\mathcal{A}_6$ model.

\subsection{Reduction to the massive \texorpdfstring{$\mathcal{A}_6$}{A6} model}\label{sub:reductionM5}

We now reduce the action \eqref{fullM5} to the lowest spherical harmonics, following \cite{Artime:2026mph}. It is useful to recall first how this works for the membrane, where the analogous step leads from the worldvolume theory to the matrix model \cite{deWit:1988wri}. The light--cone membrane is described by the embedding functions $X^I(\sigma)$ on a two--sphere, and its potential is the square of their Poisson bracket, i.e.\ of the Jacobian of two functions. Keeping only the three lowest harmonics, $X^I=X^I_ax_a$, with $x_a$ the coordinates of $\mathbf{S}^2\subset\mathbb{R}^3$, the Poisson bracket of two fields is again linear in the $x_a$. Its coefficient is precisely the commutator of the $\mathfrak{su}(2)$ matrices $X^I=X^I_a\sigma_a$, so that the reduction amounts to the replacement
\begin{eqn}\label{membranerule}
 \{X^I,X^J\}=\epsilon_{abc}X^I_bX^J_c\,x_a
 \quad\longrightarrow\quad
 -\frac\ii2\,[X^I,X^J]=\epsilon_{abc}X^I_bX^J_c\,\sigma_a .
\end{eqn}
In this way the Poisson bracket of functions becomes the commutator of matrices, and the membrane becomes the $\mathfrak{su}(2)$ matrix model.

\smallskip

The fivebrane works in the same way. We expand all the fields on the six coordinates $u_a$ of the five--sphere, in the frame of \eqref{dictionary},
\begin{eqn}\label{M5dictionary}
 X^I=X_a^Iu_a,\qquad \psi=\frac1{\sqrt2}\Theta_au_a,\qquad
 \mathsf H=\frac1{3!}H_{abc}\,du_a\wedge du_b\wedge du_c ,
\end{eqn}
so that the scalars and the fermions are linear functions, while the tensor is a constant three--form of $\mathbb{R}^6$ restricted to the sphere. The two worldvolume brackets now play the role of the Poisson bracket. As shown in Appendix~\ref{app:sphereM5}, the bracket of the fields is again a linear function, and its coefficient is the corresponding bracket \eqref{brackets} of the finite model,
\begin{eqn}\label{bracketreduction}
 \N(X^{I_1},\ldots,X^{I_5})=-6\,C_a^{I_1\ldots I_5}\,u_a,
 \quad&\longrightarrow\quad [X^{I_1},\ldots,X^{I_5}]_a=C_a^{I_1\ldots I_5},\\
 \Bh(X^I,X^J)=-6\,B_a^{IJ}\,u_a ,
 \quad&\longrightarrow\quad [X^I,X^J]_a=B_a^{IJ} .
\end{eqn}
Hence the reduction replaces the Jacobian of five functions by the five--bracket of $\mathcal{A}_6$, with structure constants $F=\epsilon/6$, and the bracket defined by the flux by the two--bracket, with structure constants $H_{abc}$. This is the fivebrane analogue of \eqref{membranerule}. The only difference is that the five--bracket is not a commutator of matrices, but the bracket of the $5$--Lie algebra $\mathcal{A}_6$, while the two--bracket is a commutator whose structure constants are the dynamical flux. This explains why the structure constants of BFSS theory are replaced by $H_{abc}$. The same rule applies to the Yukawa couplings, where one entry of the brackets is the fermion, and gives $W_H$ and $W_F$.
The background terms reduce in a similar way. The potential $U$ is quadratic, so it gives the two scalar masses, while the last term in \eqref{fullM5} gives the fermion mass $\frac{\ii\mu}{8}\Theta^TG\Theta$. The most interesting terms are the couplings to the background potentials. Collecting the residual coefficient $\mathcal Z=p\,\Omega_5/6$, the integrals of Appendix~\ref{app:sphereM5} give\footnote{Since both brackets come with the same factor, it can be absorbed in the normalization, and we choose units with $6T_5/p=1$ (see eq. \eqref{radiusphysical}).}
\begin{eqn}\label{electricreduction}
 \frac{T_5}{\mathcal Z}\int(-h\wedge c_{2,B})
 &=\frac\mu6\,\epsilon^{ijk}H_{abc}Y_a^iY_b^jY_c^k,\\
 \frac{T_5}{\mathcal Z}\int c_{5,B}
 &=\frac\mu{6\cdot5!}\,(E_5)_a^{m_1\ldots m_5}C_a^{m_1\ldots m_5} .
\end{eqn}
These are exactly the two polarization terms of \eqref{Vsquares}, with $\alpha=\frac\mu3$ and $\beta=\frac\mu6$. In particular, the relation $\alpha=2\beta$, which in Section~\ref{sec:mass} was imposed by supersymmetry, here follows from the geometry of the plane--wave, since the three--form and the six--form couple to the brane with the same strength $\mu$. Collecting all the contributions, we find
\begin{eqn}\label{actionidentity}
 \left.S_{\mathrm{M5}}\right|_{\mathrm{spherical\ modes}}=
       \mathcal Z\int dt\,L_\mu ,
\end{eqn}
i.e.\ the fivebrane action restricted to the lowest harmonics is the action \eqref{Lmass} of the massive $\mathcal{A}_6$ model, with an overall factor $\mathcal Z$ coming from the integral over the sphere.\footnote{It does not affect the classical equations, but it fixes the normalization with respect to $\hbar$ when the theory is quantized at fixed light--cone momentum \cite{Susskind:1997cw,Seiberg:1997ad,Sen:1997we}.}

\medskip

The massive $\mathcal{A}_6$ model therefore describes a finite classical sector of the plane--wave M5 with regular embeddings. Its derivation parallels the passage from the plane--wave membrane to BMN \cite{Berenstein:2002jq,Dasgupta:2002hx}, while retaining the chiral tensor and five-bracket characteristic of the M5--brane. 

\section{The BLG and the \texorpdfstring{$\mathcal A_6$}{A6} theory}\label{sec:blg}

The Bagger--Lambert--Gustavsson (BLG) theory was the first Lagrangian proposal for multiple M2--branes with the full $\mathcal N=8$ superconformal symmetry and $SO(8)$ R-symmetry expected in three dimensions \cite{Bagger:2006sk,Bagger:2007jr,Bagger:2007vi,Gustavsson:2007vu}. The novelty was that its matter fields take values in a three--algebra, namely a vector space with a totally antisymmetric trilinear bracket 
\begin{eqn}
    [e_a,e_b,e_c]=f_{abcd}e_d\,,
\end{eqn}
obeying a generalized Jacobi (fundamental) identity, in the spirit of the Nambu bracket \cite{Nambu:1973qe}.\footnote{See also \cite{Blumenhagen:2025nla} for recent developments on non--associative algebra.}
Analogously to the model described in this work, the gauge invariance and closure of the supersymmetry algebra require the structure constants $f_{abcd}$ to be totally antisymmetric, and if the metric is positive definite, the only non--abelian finite-dimensional indecomposable solution is the four-dimensional algebra $\mathcal{A}_4$, with $f_{abcd}\propto\epsilon_{abcd}$ \cite{Papadopoulos:2008sk,Gauntlett:2008uf}.\footnote{This rigidity is the three-algebra counterpart of the classification of simple $n$-Lie algebras, according to which, for every $n\geq3$, over the complex numbers the only finite-dimensional simple example is the $(n+1)$-dimensional algebra $A_{n+1}$ \cite{Ling:1993nLie,deAzcarraga:2010mr}.}



The same structures have been argued to connect M2--branes to M5--branes. When the three--algebra is taken to be the infinite-dimensional Nambu--Poisson algebra of functions on an internal three-manifold $N^3$, the BLG action describes a single M5--brane on $\mathbb R^{1,2}\times N^3$ in a large three-form background \cite{Ho:2008nn,Ho:2008ve}. In particular, the BLG gauge field is identified with the components of the self-dual two-form and part of the scalar sector. This is in line with the picture of M2--branes blowing up into an M5--brane through a fuzzy three-sphere \cite{Basu:2004ed}, and with the possible appearance of three-algebras in proposed equations of motion for a non-abelian $(2,0)$ tensor multiplet \cite{Lambert:2010wm}. In this sense, the M5--brane emerges from the M2--brane theory once the three-algebra is enlarged.

\medskip

Given the multiples similarities it is therefore natural to compare the (massive) $\mathcal{A}_6$ matrix model to the (massive) BLG theory.
In this section we show that the BLG theory with the three-algebra $\mathcal{A}_4$ sits inside the $\mathcal A_6$ model. The starting observation is purely algebraic. Fixing two entries of a five-Lie bracket $\mathcal{A}_6$ defines a three-Lie bracket $\mathcal{A}_4$ (while freezing a third one gives $\mathcal{A}_3 \simeq \mathfrak{su}(2)$). The physical point is that we realize the first of these reductions dynamically. The $\mathcal{A}_4$ algebra appears on a Coulomb branch where two transverse scalars acquire expectation values $v$ along two orthogonal internal directions (see Section ~\ref{sub:blg:branch}). At leading order in the fluctuations, the $\mathcal A_6$ action and its supersymmetry transformations then becomes exactly the reduction of the BLG action to $0+1$ dimensions, with
\begin{eqn}\label{blg:feff}
 f_{abcd}=v^2F_{abcd56}=\frac{v^2}{6}\,\epsilon_{abcd},\qquad a,b,c,d=1,\ldots,4\,.
\end{eqn}
Interestingly, the eighth BLG scalar, which is not visible among the scalars $X^I$, is supplied by the chiral tensor $H_{abc}$, and the BLG gauge field arises from the tensor components with one leg along the expanded directions. This is the analogue of the identification of \cite{Ho:2008nn,Ho:2008ve} recalled above. Moreover, only an $SO(7)$ subgroup of the R-symmetry is manifest in the parent model, while the reduced theory is $SO(8)$-covariant.
Finally, we extend this relation to the massive case. Here the situation is different, since the plane--wave masses lift the flat directions. In Section~\ref{sub:blg:massive} we show that the relation survives, but around the $\mathcal{S}$ vacuum rather than on a Coulomb branch. We show that in the BLG frame the spherical vacuum $\mathcal{S}$ is a fuzzy three--sphere times a disc of the same radius, and around it the massive $\mathcal{A}_6$ model reproduces the mass--deformed BLG theory \cite{Hosomichi:2008qk,Gomis:2008cv} around its fuzzy three--sphere vacuum. 

\subsection{BLG on the $\mathcal{A}_6$ Coulomb branch}\label{sub:blg:branch}

The first step towards this identification is purely algebraic. The structure constants $F_{abcdef}$ define the simple $5$--Lie algebra $\mathcal{A}_6$,
\begin{eqn}\label{blg:fivebracket}
[e_a,e_b,e_c,e_d,e_e]=F_{abcdef}\,e_f .
\end{eqn}
In our case, fixing the last two entries to $e_5$ and $e_6$ and restricting to the remaining four generators, we obtain
\begin{eqn}\label{blg:A6toA4}
 [e_a,e_b,e_c,e_5,e_6]=\frac16\,\epsilon_{abcd}\,e_d ,\qquad a,b,c,d=1,\ldots,4,
\end{eqn}
which is the three--algebra $\mathcal{A}_4$ of BLG theory. Fixing a further entry to $e_4$ gives a Lie bracket on the first three generators, i.e.\ $\mathcal{A}_3\simeq\mathfrak{su}(2)$. We therefore have the trivial chain
\begin{eqn}\label{blg:chain}
 \mathcal{A}_6\ \xrightarrow{\ e_5,\,e_6\ }\ \mathcal{A}_4\ \xrightarrow{\ e_4\ }\ \mathcal{A}_3\simeq\mathfrak{su}(2) .
\end{eqn}
This chain has a simple geometric meaning on the spherical functions. The unit five--sphere can be seen as a family of three--spheres over the disc $D^2$, through the projection $u\mapsto(u_5,u_6)$. Fixing $u_5$ and $u_6$ in the Nambu bracket of $\mathbf{S}^5$, we are left with a bracket of three functions on the $\mathbf{S}^3$ fibres, which is, up to normalization, the Nambu bracket of $\mathbf{S}^3$. In the same way, the second arrow in \eqref{blg:chain} is the reduction used in \cite{Mukhi:2008ux} to obtain D2--branes from BLG theory.

\medskip

We now realize the first arrow dynamically. Consider the configuration
\begin{eqn}\label{blg:vev}
 X_a^8=v\,\delta_{a5},\qquad X_a^9=v\,\delta_{a6},\qquad
 X_a^I=0\  ,\qquad H=\Theta=0 .
\end{eqn}
Only two scalar vectors are nonzero, so the five--bracket vanishes, $C=0$, and the two--bracket vanishes because $H=0$. Hence, in the massless model, \eqref{blg:vev} has zero energy, satisfies the Gauss' law \eqref{gauss}, and preserves all sixteen dynamical supersymmetries (since $\delta\Theta$ in \eqref{susy} vanishes). In other words, it is a flat direction representing a point on a Coulomb branch. It breaks the gauge group $SO(6)\to SO(4)$, and the target symmetry $SO(9)\to SO(7)\times SO(2)_D$, where $SO(2)_D$ rotates the target plane $(8,9)$ together with the internal plane $(5,6)$. Through the dictionary \eqref{M5dictionary}, the corresponding fivebrane is then described by $X^8+\ii X^9=v\,(u_5+\ii u_6)$. It is just the projection of $\mathbf{S}^5$ on the disc where the fivebrane extends along a disc of radius $v$ in the $(8,9)$ plane, while its $\mathbf{S}^3$ fibres are collapsed.

\smallskip

We now organize the fluctuations around \eqref{blg:vev} according to the reduction \eqref{blg:A6toA4}. The scalars $X^I_a$ with $I\le7$ and $a\le4$ are kept as they are because they will not play a different role than before. The twenty components of the tensor are then decomposed under $SO(4)$ as $\mathbf{20}\to\mathbf4\oplus\mathbf4\oplus\mathbf6\oplus\mathbf6$, and we parametrize them as
\begin{eqn}\label{blg:tensorsplit}
 H_{abc}=f_{abcd}\,X^\star_d,\qquad
 H_{ab5}=\frac1v\,\tilde A_{1,ab},\qquad
 H_{ab6}=\frac1v\,\tilde A_{2,ab},\qquad
 H_{a56}=u_a\,.
\end{eqn}
As known the components $H_{abc}$ are then dual to a vector $X^\star_d$, which will play the role of the eighth scalar of BLG theory (while the two antisymmetric matrices $\tilde A_{1,2}$ will become its gauge field). It is therefore convenient to collect the eight scalars as
\begin{eqn}\label{blg:eightscalars}
 X^{\hat I}_a=\bigl(X^1_a,\ldots,X^7_a,X^\star_a\bigr) .
\end{eqn}
The nine gauge generators broken by \eqref{blg:vev}, i.e.\ $A_{a5}$, $A_{a6}$ and $A_{56}$, can be used to set 
\begin{eqn}
    X^8_a=X^9_a=0 \qquad \text{and} \quad  X^8_6=X^9_5\,,
\end{eqn}
while their Gauss constraints fix their momenta. The other fluctuations, namely $X^I_5$, $X^I_6$, the moduli of the disc and the fermions $\Theta_5$, $\Theta_6$, form a separate sector, which we will show that decouples at leading order.
Finally, before proceeding we have to specify in which sense the reduced theory is the leading order. The massless model has no dimensionful parameter. Hence the value of $v$ itself has no meaning per se, and the only parameter is the size of the fluctuations relative to it,
\begin{eqn}\label{blg:delta}
 \delta\equiv \frac{|X^{\hat I}|}{v} ,
\end{eqn}
in terms of which the reduced theory is the leading order of an expansion for small $\delta$.

\subsection{The reduced action}\label{sub:blg:action}

Inserting \eqref{blg:vev} and \eqref{blg:tensorsplit} into \eqref{Lzero}, and imposing the (broken) Gauss constraints, we find at leading order in $\delta$
\begin{eqn}\label{blg:action}
 L_{\rm BLG}={}&\frac12\bigl(D_tX^{\hat I}\bigr)^2
 +\frac{3}{2v^2}\,\epsilon_{abcd}\,\tilde A_{1,ab}\,D_t\tilde A_{2,cd}
 -\frac12\sum_{i=1,2}\bigl(\tilde A_iX^{\hat I}\bigr)^2
 -\frac1{12}\bigl[X^{\hat I},X^{\hat J},X^{\hat K}\bigr]^2\\
 &+\frac\ii2\Theta^TD_t\Theta+\frac\ii2\Theta^T W_{\rm BLG}\Theta
\end{eqn}
with all indices summed, and
\begin{eqn}\label{blg:bracket}
 [X,Y,Z]_a=f_{abcd}X_bY_cZ_d .
\end{eqn}
The fermion matrix is\footnote{Here we introduced the common eight real $16\times16$ matrices
\begin{eqn}\label{blg:hatgamma}
 \hat\Gamma_I=\gamma_I\,\qquad \hat\Gamma_\star=J,\qquad
 \hat\Gamma_{\hat I}^T\hat\Gamma_{\hat J}+\hat\Gamma_{\hat J}^T\hat\Gamma_{\hat I}=2\delta_{\hat I\hat J} \quad \text{with} \quad \Gamma_{\hat I\hat J}=\hat\Gamma^T_{[\hat I}\hat\Gamma_{\hat J]}\,.
\end{eqn}}
\begin{eqn}\label{blg:W}
 (W_{\rm BLG})_{ab}=\tilde A_{1,ab}\gamma_8+\tilde A_{2,ab}\gamma_9
 -\frac12f_{abcd}X^{\hat I}_cX^{\hat J}_d\,J\,\Gamma_{\hat I\hat J},
 \qquad \text{with} \quad J\equiv\gamma_{89}\,,
\end{eqn}
Since $J$ commutes with $\gamma_{I}$ and $J^2=-1$, the matrices $\Gamma_{\hat I\hat J}$ generate $\mathfrak{so}(8)$ and commute with $\gamma_8$ and $\gamma_9$. The sixteen components of $\Theta_a$ therefore decompose as $\mathbf 8_c\otimes\mathbf 2$ under $SO(8)\times SO(2)_D$, with $\gamma_8,\gamma_9$ acting on the two-dimensional factor. This is precisely the BLG spinor reduced on the two spatial worldvolume directions.

\smallskip

The computation is due to a direct substitution and for completeness we list the origins of the various terms of \eqref{blg:action}. 

\begin{itemize}
\item \emph{Chern--Simons term.} The tensor kinetic term $\frac12HFD_tH$ restricted to $(\tilde A_1,\tilde A_2)$ is the first-order term in \eqref{blg:action}.
The coefficient is exactly that of the BLG term reduced to $0+1$ dimensions. 

\item \emph{The eighth scalar.} The first--order tensor term pairs each triple with its complement, so $X^\star$ is paired with $u$. The two--bracket on the disc gives a quadratic term in $u$, and eliminating $u$ through its equation of motion gives the kinetic term $\frac12(D_tX^\star)^2$.

\item \emph{Covariant derivatives along the M2.} With $D_iX\equiv-\tilde A_iX$, the components $B^{8I}$ and $B^{9I}$ give the covariant derivative term along the M2. The corresponding terms for $X^\star$ instead are generated by the broken Gauss constraints, which fix $P^8_a$ and $P^9_a$.

\item \emph{Potential and Yukawa couplings. } The components $C^{IJK89}_a$ and $B^{IJ}_a$ combine into the $SO(8)$-invariant BLG potential. The Yukawa couplings combine in the same way. The term $W_F$ provides $\frac12fXX\,\gamma_{IJ}J$, while $W_H$ provides the rest. 
\end{itemize}

\medskip

To complete the discussion we will write explicitly the supersymmetry transformations \eqref{susy0}, using the ansatz \eqref{blg:vev} and \eqref{blg:tensorsplit}. The sixteen dynamical supersymmetries reduce to
\begin{eqn}\label{blg:susy}
 \delta X^{  I}_a&=-\ii\,\epsilon^T \hat\Gamma_{  I}\Theta_a,\\
 \delta\tilde A_{i,ab}&=\ii\,f_{abcd}X^{  I}_c\,\epsilon^T \hat\Gamma_{  I}\gamma_{7+i}J\,\Theta_d,\\
 \delta\Theta_a&=\Bigl[D_tX^{  I}_a\, \hat\Gamma^T_{  I}
 +(D_iX^{  I})_a\,\gamma_{7+i} \hat\Gamma^T_{  I}
 +\frac16[X^{  I},X^{  J},X^{  K}]_a\,J\, \hat\Gamma_{  I  J  K}\Bigr]\epsilon ,
\end{eqn}
where we removed the hat on the indices only for better readability. 
The different variation of the fields are obtained in complete analogies to how to obtain the Lagrangian \eqref{blg:action}.

\subsubsection*{Interpretation and comments}\label{sub:blg:comments}

The reduction reverses the logic of \cite{Ho:2008nn,Ho:2008ve}. There, the Nambu--Poisson algebra on $N^3$ turns BLG into an M5 on $\mathbb R^{1,2}\times N^3$. Here, the M5 described by the $\mathcal A_6$ model is stretched along two transverse directions, and its remaining three worldvolume directions are the $\mathbf{S}^3$ fibres of $\mathbf{S}^5 \to \mathbf{D}^2$. The truncation of $\mathbf{S}^5 $ to linear harmonics then induces the truncation of $\mathbf{S}^3$ to linear harmonics, which is $\mathcal{A}_4$. The dictionary between the two descriptions is
\begin{eqn}\label{blg:dictionary}
 \underbrace{H_{ab5},\ H_{ab6}}_{b_{\mu\dot\nu}}\ \longleftrightarrow\ \tilde A_{1,2},\qquad
 \underbrace{H_{abc}}_{b_{\dot\mu\dot\nu}}\ \longleftrightarrow\ X^\star,\qquad
 X^{1,\ldots,7}\ \longleftrightarrow\ X^{1,\ldots,7},
\end{eqn}
in the notation of \cite{Ho:2008nn,Ho:2008ve}, where dotted indices refer to $N^3$ and undotted ones to the M2 worldvolume. In other words, the three-bracket is the five-bracket with two entries frozen along the M2 worldvolume and the $\mathcal A_6$ model generalizes BLG in the same way that the M5 generalizes the M2. The theory is then reduced to quantum mechanics because no momentum modes along the disc survive the truncation!
In this sense, the $SO(8)$ of BLG is broken to the $SO(7)$ of D2 super Yang--Mills \cite{Mukhi:2008ux}, while here the situation is reversed. Only $SO(7)$ is manifest, and the eighth direction, being part of the chiral tensor, mixes with the transverse scalars only in the limit. 

A similar enhancement was also found in \cite{Bandos:2008fr}, where the infinite--tension limit of an M5--brane on $\mathbb R^2\times M_3$ gives the Carrollian limit of BLG theory with the Nambu--Poisson algebra of functions on $M_3$, with the eighth scalar again supplied by the two--form. This limit suppresses only the derivatives along $\mathbb R^2$, while the couplings to the gauge field survive. Hence, for $M_3=\mathbf S^3$, keeping the four coordinate harmonics and restricting to fields homogeneous on $\mathbb R^2$ gives again \eqref{blg:action}. Here, instead, the same theory arises dynamically, at leading order around the two--scalar Coulomb branch \eqref{blg:vev}, and with the dynamical coupling $f_{abcd}=v^2\epsilon_{abcd}/6$.\footnote{We thank Ralph Blumenhagen for pointing this out to us.}
Within the reduced theory, a further expectation value 
\begin{eqn}
    \langle X^\star_a\rangle= w \,\delta_{a4}\,,
\end{eqn}
realizes the $\mathcal{A}_4 \,\,\to \mathfrak{su}(2)$ expressed graphically in \eqref{blg:chain}. By \eqref{blg:tensorsplit}, this is a tensor background $H \propto e^{123}$, which is the normal form \eqref{tensorNFform} with $b=0$. This suggests that the $\mathfrak{su}(2)$ sector obtained from the large-flux scaling \eqref{largefluxscaling} can be read in the BLG frame as the reduction of BLG theory to D2--branes of \cite{Mukhi:2008ux}.

\medskip

We conclude that \emph{the BLG theory with three-algebra $\mathcal{A}_4$, reduced to $0+1$ dimensions, is the low-energy theory on the two-scalar Coulomb branch of the massless $\mathcal A_6$ model.}

\subsection{Massive $\mathcal{A}_6$ model and massive BLG}\label{sub:blg:massive}

We now ask whether the analysis of the previous sections extends to $\mu\neq0$. Mass deformations of BLG theory have been studied before. The best known one preserves sixteen supersymmetries, breaks $SO(8) \to SO(4)\times SO(4)$, gives the same mass $m$ to all eight scalars and contains a quartic flux term \cite{Hosomichi:2008qk,Gomis:2008cv}. It is indeed this flux term which allows massive BLG theory to admit a fuzzy three-sphere vacuum \cite{Hosomichi:2008qk}. 

\medskip
 
In the massless case three facts make the reduction supersymmetric: \eqref{blg:vev} is a flat direction, it preserves all sixteen supersymmetries, and the disc sector $X^I_{5,6}$, $\Theta_{5,6}$ decouples at leading order in $\delta$. For $\mu\neq0$ these facts are not anymore true. Indeed, the disc lies in the massive six-plane and costs $V=\beta^2v^2$. As explained in Section~\ref{sec:vacua}, this has no supersymmetric vacuum of this rank. Additionally, the plane--wave mass term induces another scale to which compare the perturbative expansion. For masses of the order of the BLG interactions, $\mu\sim v^2|X|^2$, the cross terms with the disc are as large as the terms kept, and the above perturbation expansion cannot be trusted anymore. 

\medskip

To check if a similar equivalence may be expressed in different terms we will then compare the two Lagrangians and aim to understand how a similar analysis can be done.
Proceeding in this direct way, we insert the ansatz into \eqref{Lmass}. The polarization term with $H_{abc}=f_{abcd}X^\star_d$ gives straightforwardly\footnote{We now need to respect also the splitting given by the plane--wave background. We then split the BLG scalars as $A,\ldots\in\{1,2,3,\star\}$ and $A',\ldots\in\{4,\ldots,7\}$.}
\begin{eqn}\label{blgm:reduced}
 L^{(\mu)}_{\rm red}=L_{\rm BLG}-\frac{\alpha^2}2\bigl(X^i_a\bigr)^2-\frac{\beta^2}2\bigl(X^{A'}_a\bigr)^2
 +\frac\beta4\Bigl(\epsilon^{ABCD}+\epsilon^{A'B'C'D'}\Bigr)f_{abcd}X_aX_bX_cX_d+\frac{\ii\mu}8\Theta_a^TG\Theta_a\,.
\end{eqn}
Now, we need to compare this Lagrangian to the known (0+1 reduction of the) Lagrangian of massive BLG theory \cite{Hosomichi:2008qk,Gomis:2008cv}
\begin{eqn}\label{blgm:HLL}
 L_{\rm mBLG}=L_{\rm BLG}-\frac{m^2}2\bigl(X^{\hat I}_a\bigr)^2
 +\frac m6\Bigl(\epsilon^{ABCD}+\epsilon^{A'B'C'D'}\Bigr)f_{abcd}X_aX_bX_cX_d
 +\frac{\ii m}2\Theta_a^TG\Theta_a ,
\end{eqn}
We can already note that the two Lagrangians do not match. However, the quartic and fermion mass terms coincide for $m=\frac32\beta=\frac\mu4$. Unfortunately, the scalar masses still do not match due to the different coefficients and especially because in \eqref{blgm:HLL} the mass of $X^\star$ is precisely 0 (instead of $m$). 
To have intuition about why this happens, it is very important to know how supersymmetry closes in the massive $\mathcal{A}_6$ model. To be precise, the reason follows from \eqref{susy:parameter}: both theories satisfy it with the same $m_F=-m$, but the massive BLG theory has $k=0$, while in the massive $\mathcal A_6$ it is forced $\alpha=2\beta$, i.e. $k \neq 0$. 

\smallskip

Indeed the obstruction is not in the algebra but in the background. In the massless case, BLG theory is the low-energy theory around a supersymmetric zero-energy configuration, and for $\mu\neq0$ the disc \eqref{blg:vev} is no longer one. Hence, \emph{if} a massive BLG theory sits inside $\mathcal A_6$, it must be found by expanding around one of the classical vacua of Section~\ref{sec:vacua}. Among them, $\mathcal S$ is the main candidate.\footnote{In the $Z=0$ vacua the five-bracket vanishes identically and only the $\mathfrak{su}(2)$ structure obtained by freezing three entries survives.} This is the same mechanism that occurs in the massive BLG theory itself: the mass deformation lifts the flat directions and replaces them by polarized vacua, i.e.\ fuzzy spheres supported by the flux \cite{Myers:1999ps,Berenstein:2002jq,Hosomichi:2008qk}. Therefore, the natural question is not whether the Coulomb branch survives, but whether $\mathcal S$ looks like the fuzzy three-sphere of the massive BLG theory once two of its directions are frozen.

In the BLG frame, $\mathcal S$ is the disc \eqref{blg:vev} times a fuzzy three-sphere $X^{A'}_a=R\,\delta_{a,A'-3}$, so that the squares in \eqref{Vsquares} split as
\begin{eqn}\label{blgm:split}
 V=\underbrace{2R^2\Bigl(\frac{R^2v^2}6-\beta\Bigr)^2}_{a\leq4}+\underbrace{v^2\Bigl(\frac{R^4}6-\beta\Bigr)^2}_{a=5,6}\,,
\end{eqn}
with $f_{1234}=v^2/6$. The first term is the fuzzy three-sphere potential of the massive BLG theory with the shift $\beta=\frac23m$, and vanishes for $R^2v^2=\mu$. The second fixes 
\begin{eqn}
     R=v=\mu^{1/4}\,,
\end{eqn}
which is the vacuum $\mathcal S$. What is relevant in this analysis is that the BLG sector alone vanishes on the whole curve $R^2v^2=\mu$, which collects the fuzzy three-sphere vacua of BLG theory for all values of the coupling $f_{1234}=v^2/6$. Since here $v$ is dynamical, these would form a flat direction. The disc sector then lifts it exactly to selects the isolated point $R=v=\mu^{1/4}$, i.e.\ $\mathcal S$. Indeed, the quartic coupling of \eqref{blgm:reduced} is shared as $\frac\beta4=\frac\beta6+\frac\beta{12}$ between the two sectors. In other words, the fuzzy three-sphere exists only in the complete model: the BLG sector imposes only its shape, while the disc sector its size!\footnote{It would be interesting if a similar analysis can be done to the $\mathcal{R}$ vacua in order to select only one of the possible radius inside $\mathbf{B}^3$.}

\smallskip

 This becomes even clearer if we study the fluctuations around $\mathcal{S}$ as in Section \ref{ss:sphere}. As explained above, the main problem of this comparison comes from the fact that for massive BLG theory $\epsilon$ is a constant spinor ($k=0$).\footnote{This means that $Q$ acting of a state of energy $E$ returns a vector with the same energy. In other words, this is why its spectrum around the fuzzy three-sphere is fully degenerate and bosons and fermions are all at $2m$.} So, to compare the different fluctuations we need to remove the time dependence of $\epsilon(t)$. By \eqref{susy} this can easily done by rotating back the fermions with $G$, in order to have a constant $\epsilon$ in this frame.  This clearly leaves the bosons unchanged and moves the fermions of $\mathcal S$ to all the same mass $\frac\mu2$ (as one would expect in a reference that follows the fluxes). Taking the results from \eqref{spherespectrum}, we find
\begin{eqn}\label{blgm:spectra}
 \begin{array}{l|c|c}
  & \quad \quad \text{        mBLG      } \quad \quad  &  \mathcal S\\ \hline
  \text{fermions} & 32\ \text{at}\ \tfrac\mu2 & 32+16\ \text{at}\ \tfrac\mu2\\
  X^{A}\,,\ \tilde A_{1,2} & 22\ \text{at}\ \tfrac\mu2 & 22+6\ \text{at}\ \tfrac\mu2\\
  \text{shape modes} & 10\ \text{at}\ \tfrac\mu2 & 20\ \text{at}\ \tfrac\mu3,\ 1\ \text{at}\ \tfrac{2\mu}3
 \end{array}
\end{eqn}
As explained above, and confirming it in this analysis, the extra $16+6$ modes are only due to the disc sector. Hence, the only difference is in the shape modes, which in $\mathcal A_6$ are the $\mathbf{20}'\oplus\mathbf 1$ of $Z$ rather than the $\mathbf 9\oplus\mathbf 1$ of $X^{A'}$ (modulo $SO(4)$). Physically speaking this is clear because in BLG theory $f_{abcd}$ is fixed, while here $f_{abcd}=v^2\epsilon_{abcd}/6$ with $v$ dynamical, and at $\mathcal S$ ($R=v$) the disc and the three-sphere mix.

\medskip

We conclude that \emph{the massive $\mathcal A_6$ model reproduces the massive BLG theory around its fuzzy three-sphere vacuum, up to the coupling between the shape modes and the dynamical BLG coupling}! Without masses, BLG theory lives on a Coulomb branch where the M5--brane is a collapsed disc, while with masses it lives around the round M5--brane itself, in line with \cite{Gomis:2008vc,Gustavsson:2009qd} (see also \cite{Lin:2005nh}). This again supports that $\mathcal S$ represents a single spherical M5--brane.

\section{Discussion and future directions}\label{sec:conclusion}

In this work we construct the maximally supersymmetric plane--wave deformation of the $\mathcal{A}_6$ matrix model for a spherical M5--brane, which we call plane--wave or massive $\mathcal{A}_6$ model.

After proving that the model preserves sixteen dynamical and sixteen kinematical supersymmetries, and writing down explicitly its equations of motion, we study its vacua. We derive the complete list of zero--energy configurations. The first one is a stable round five--sphere $\mathbf{S}^5\subset\mathbb{R}^6$ with vanishing $H_{abc}(t)$ tensor, which we denote by $\mathcal{S}$. The other solutions have $Z=0$ and are supported by the dynamical $H_{abc}(t)$ tensor, which can now balance the size of the brane. In this way we find continuous families in the three--plane $\mathbb{R}^3$, where a larger radius is compensated by a smaller flux. These solutions leave free the components of the tensor with legs in both internal triples, which can be nonzero without breaking any supersymmetry. The two resulting families, denoted by $\mathcal{R}_{\rm I}$ and $\mathcal{R}_{\rm II}$, have respectively an $SO(3)$ stabilizer and a trivial one. We show that the image of these families is a three--ball $\mathbf{B}^3$, and \emph{not} a two--sphere $\mathbf{S}^2$, as one could expect for a membrane. Indeed, the fivebrane is collapsed onto the three--plane, and its five--sphere is projected onto a ball foliated by concentric two--spheres. The distinction becomes clearer by studying the algebra induced by the $H_{abc}(t)$ tensor, which defines an $\mathfrak{su}(2)$ Poisson bracket whose leaves are exactly these two--spheres. On $\mathcal{R}_{\rm I}$ the two internal triples generate two commuting algebras, $\mathfrak{su}(2)\oplus\mathfrak{su}(2)$, or $\mathfrak{su}(2)$ and an abelian factor when the second flux vanishes. On $\mathcal{R}_{\rm II}$, instead, the tensor mixes the two triples. However, after a simple change of basis the algebra again splits into $\mathfrak{su}(2)\oplus\mathfrak{su}(2)$, but now the occupied vectors have nonzero components along both factors. In other words, the $\mathfrak{su}(2)$ generated by the scalars is embedded diagonally, which explains algebraically why the stabilizer is trivial. Due to the appearance of these $\mathfrak{su}(2)$ algebras it is natural to ask whether this suggests an underlying BMN structure. Indeed, in an appropriate large--flux limit, the Lagrangian of the massive $\mathcal{A}_6$ model reduces to the one of the $\mathfrak{su}(2)$ BMN model, whose structure constants are given by the tensor and in general generate two $\mathfrak{su}(2)$ factors.

Finally, we find a collapsed cone of pure flux. In this case all the scalars vanish, so the fivebrane has zero size, but the chiral tensor can still be nonzero. Since the worldvolume three--form of the M5--brane carries dissolved membrane charge, this solution may describe membrane charge stored on a collapsed fivebrane following the $\mathcal{R}$ solution. Its energy vanishes because the flux couples to the scalars only through the size of the brane, so once the fivebrane has collapsed the flux no longer costs energy. This is the fivebrane counterpart of a D2--brane with internal flux on a vanishing cycle, where the dissolved D0--charge survives the collapse of the cycle. The Gauss' law allows only two components of the tensor, and the residual $SO(6)$ acts on them as a quarter rotation, as electric--magnetic duality acts on a pair of charges. Hence this solution is described by $\mathbb{C}/\mathbb{Z}_4$, and the full $SO(6)$ is restored only at its apex, where the flux vanishes, which corresponds to the $X=0$ solution connected to the $\mathcal{S}$ one.

For completeness, we also consider more general round configurations in the two planes, and we find a pair of new stationary solutions with positive energy. They are unstable, since they sit on top of the barrier which separates the $\mathcal{R}$ and $\mathcal{S}$ solutions from the collapsed configuration $X=0$ with and without fluxes, respectively. We construct the corresponding instantons of the Euclidean theory, and we compute their action, which describes the tunnelling between the collapsed and the expanded vacua. Interestingly, the unstable solution associated to $\mathcal{R}$ is sustained by a tensor flux which grows linearly in time.

\smallskip

We then study the fluctuations around each supersymmetric vacuum. Their frequencies are organized into multiplets of $\mathfrak{su}(2|4)$, in which bosons and fermions energies differ by multiples of $\mu/12$, since the supercharges depend on time \cite{Dasgupta:2002hx}. As usual in plane--wave backgrounds, the numbers of bosonic and fermionic oscillators differ, but the zero--point energies still cancel. Most importantly, around $\mathcal{S}$ the fluctuations form the $\ell=1$ multiplet of the spherical transverse fivebrane, whose $\ell=0$ multiplet describes the trivial vacuum of BMN theory. Instead, around the flux--supported families they reproduce the multiplets of BMN theory around a fuzzy sphere. Hence $\mathcal{S}$ realizes classically the spherical fivebrane that in BMN theory was identified, at large $N$ and strong coupling, through the protected spectrum above $X=0$ \cite{Maldacena:2002rb} (see also \cite{Asano:2017xiy,Asano:2017nxw}).

This comparison \emph{may} suggest a simple relation between the two models. In BMN theory the degrees of freedom are matrices, which regularize the membrane. The membranes are classical vacua, the fuzzy spheres, while the fivebrane appears only at large $N$ and strong coupling, through the protected states above the trivial vacuum. In the BMN picture the states are composite and the lowest state of the $\ell=1$ multiplet is made of two oscillators of the matrices. In the massive $\mathcal{A}_6$ model the degrees of freedom are those of the fivebrane, and the roles are reversed. In this work the fivebrane is a classical vacuum, and the same multiplet is made of its elementary fluctuations. The membranes, instead, may appear only indirectly through the flux--supported vacua, which carry membrane charge. Since the energies of the protected states do not depend on the coupling, this comparison is meaningful even though one model is quantum and the other is classical. It may therefore suggest to regard the two models as complementary descriptions of the spherical branes of the plane--wave, one natural for membranes and the other for fivebranes, in the spirit of the electric--magnetic relation between the two branes in eleven dimensions.

\smallskip

As an independent confirmation, we derive the light--cone action of the M5--brane in the maximally supersymmetric plane--wave, using the Hamiltonian formulation of \cite{Bergshoeff:1998vx} and its light--cone gauge \cite{Bandos:2008fr}, and we reduce it to the lowest spherical harmonics. Extending the flat--space truncation of \cite{Artime:2026mph}, we show that the result is exactly the massive $\mathcal{A}_6$ model, including the dynamics of the self--dual tensor and the masses of bosons and fermions, and that it describes a consistent classical sector of a single spherical M5--brane. This derivation adds two things to the supersymmetric construction. On the one hand, it gives each term of the model a spacetime interpretation, which the Noether construction does not provide. On the other hand, since the truncation is consistent, every classical solution of the model is a solution of the full fivebrane. In other words, $\mathcal{S}$ is the classical M5 giant graviton of the plane--wave, truncated to its lowest harmonics, in the same way in which the fuzzy sphere of the $\mathfrak{su}(2)$ BMN model is the lowest truncation of the membrane giant graviton.

\smallskip

Finally, we establish the relation between the (massive) $\mathcal{A}_6$ model and (massive) BLG theory \cite{Bagger:2006sk,Bagger:2007jr,Bagger:2007vi,Gustavsson:2007vu}. In the massless case, we prove that BLG theory with the three--algebra $\mathcal{A}_4$, reduced to $0+1$ dimensions, is the low--energy theory on a Coulomb branch where two scalars acquire an expectation value $v$. On this branch the fivebrane collapses to a disc, and the three--bracket is obtained by freezing two entries of the five--bracket along it. Hence its structure constants $f_{abcd}=v^2\epsilon_{abcd}/6$ are not fixed numbers, but the BLG coupling becomes a dynamical modulus. Moreover, the eighth scalar and the gauge field of BLG theory are supplied by the chiral tensor, as in the identification of the BLG fields with the M5 two--form of \cite{Ho:2008nn,Ho:2008ve, Bandos:2008fr}, so that the $SO(7)$ symmetry of the parent model is enhanced to $SO(8)$. In the massive case the masses lift the Coulomb branch, and the relation appears instead around $\mathcal{S}$, which in the BLG frame is a fuzzy three--sphere times a disc of the same radius. Around it, the massive $\mathcal{A}_6$ model reproduces the mass--deformed BLG theory around its fuzzy three--sphere vacuum \cite{Hosomichi:2008qk,Gomis:2008cv}, except for the shape modes of the three--sphere, which couple to the radius of the disc, i.e.\ to the dynamical BLG coupling. Physically speaking, without masses the two M2--branes live on the collapsed fivebrane, while with masses they are polarized into the round fivebrane itself, in line with the interpretation of the fuzzy three--sphere vacua as M5--branes \cite{Gomis:2008vc}.

\medskip

This work is just a first step in the study of these models, and many directions remain open.
Since the fluctuations around each vacuum are gapped and the expansion is controlled by $\mu$, the quantum theory can be studied semiclassically without a large $N$ limit. The natural next step is therefore its quantization, and the computation of a supersymmetric index, as has been done for the finite--$N$ BMN model and in connection with ABJM \cite{Aharony:2008ug, Chang:2024lkw, Chang:2026yeg, Komatsu:2026ano, Komatsu:2026tpi}. In our case this requires a careful treatment of the tensor sector, since the flux--supported families and the cone of pure flux are flat directions, which give a continuum of zero--energy states unless the flux is quantized. The same quantization may also clarify the nature of the flux--supported vacua. Their spectrum coincides with that of BMN theory around a fuzzy sphere, but classically their image is a three--ball, since no single symplectic leaf of the $\mathfrak{su}(2)$ bracket is selected. If the quantized flux reproduces the relation between radius and representation of BMN theory \cite{Berenstein:2002jq,Dasgupta:2002hx}, $\mathcal{R}$ would describe a single spherical fuzzy membrane.

The main open problem is the absence of a large $N$ limit, which in matrix theory is needed for the membrane and fivebrane charges to appear in the supersymmetry algebra. Since $\mathcal{S}$ reproduces the $\ell=1$ level of the multiplets of the spherical fivebrane, a complete description should include the whole tower of harmonics. Alternatively, the embedding $\mathcal{A}_4\subset\mathcal{A}_6$ may suggest an extension analogous to the passage from BLG to ABJM theory \cite{Aharony:2008ug,Bagger:2008se}, with less supersymmetry and a level, which here is already visible in the dynamical BLG coupling. Such an extension would also be a natural starting point for a holographic description, in the spirit of the AdS/CFT correspondence \cite{Maldacena:1997re,Gubser:1998bc,Witten:1998qj,Aharony:1999ti}, would have to face the absence of a Lagrangian description of the interacting $(2,0)$ theory.

An orthogonal direction is to include brackets of higher order. In the construction of \cite{Artime:2026fcr}, closure on the scalars restricts the possible orders of the brackets, and besides the two-- and five--brackets the allowed orders include six and nine. The authors conjectured that these might describe the Kaluza--Klein monopole and the Ho\v{r}ava--Witten domain wall, respectively. A possible route would be to opportunely replace the $\mathcal{A}_6$ action by the simple $6$-- and $9$--Lie algebras $\mathcal{A}_7$ and $\mathcal{A}_{10}$, and to recover the known models by freezing entries of the brackets, as we did to obtain BLG theory.

\smallskip

We hope that the results presented in this work offer new insights into the description of fivebranes, and that they serve as a useful basis for future explorations.


\medskip

\paragraph{Acknowledgements.}
The author thanks Manuel Artime, Ralph Blumenhagen, Jarod Hattab, Dieter L\"ust and Max Wiesner for illuminating discussions and for useful comments on the manuscript. This work is supported by the Deutsche Forschungsgemeinschaft through the Emmy Noether 
programme -- 557478919, under Germany's Excellence Strategy EXC 2121 ``Quantum Universe'' -- 390833306, through the German--Israeli Project Cooperation (DIP) grant ``Holography and the Swampland'', and through the Collaborative Research Center CRC 1624 ``Higher Structures, Moduli Spaces and Integrability''. The author is also grateful to Karoline Regauer 
for her continuous encouragement and support.


\appendix

\section{Details on the supersymmetry transformation}\label{app:susy}
\subsection{Dirac brackets and the Gauss' law}\label{app:details}

In this Appendix we derive the brackets and the Gauss' law used in Section~\ref{sec:mass}. Only the first--order terms of the action matter for this purpose, and since the mass deformation does not modify them, we can work with the massless action \eqref{Lzero}.

\smallskip

The kinetic term of the tensor is first order in time, as is typical for a self--dual field \cite{Henneaux:1988gg,Bergshoeff:1998vx,Bandos:2008fr}. This means that the momenta of $H$ are not independent variables, but are fixed by $H$ itself. These relations are constraints on phase space, and to take them into account we have to replace the Poisson brackets by Dirac brackets \cite{Henneaux:1992ig,Faddeev:1988qp}. The same happens for the fermions, whose kinetic term is also first order.

Since $H_{abc}$ is totally antisymmetric, it has twenty independent components, which we label by the triples $p=\{abc\}$, with $H_p=H_{abc}$. 
In the kinetic term each unordered triple appears $3!$ times in both $H_{abc}$ and $H_{def}$. Using the normalized notation $F_{abcdef} \to 6F_{qp}=\epsilon_{qp}$ we obtain
\begin{eqn}\label{tensorconstraints}
 \frac12H_{abc}F_{abcdef}\dot H_{def}=3\,H_pF_{pq}\dot H_q ,
\end{eqn}
so that the momentum conjugate to $H_p$ is $\pi_p=-3F_{pq}H_q$. This gives the twenty constraints
\begin{eqn}
 \chi_p=\pi_p+3F_{pq}H_q=0 .
\end{eqn}
Using $\{H_p,\pi_q\}=\delta_{pq}$, their brackets are $C_{pq}\equiv\{\chi_p,\chi_q\}=6F_{pq}$. This matrix is invertible, $C^{-1}=-\frac16 F$, so the constraints are of second class and can be removed with the Dirac prescription,
\begin{eqn}\label{diracderivation}
 \{A,B\}_D=\{A,B\}-\{A,\chi_p\}(C^{-1})_{pq}\{\chi_q,B\} .
\end{eqn}
Since $\{H_p,\chi_q\}=\delta_{pq}$, we obtain
\begin{eqn}
 \{H_p,H_q\}_D=-\frac16 F_{pq} .
\end{eqn}
Hence each component of the tensor is conjugate to the component of the complementary triple, for instance $\{H_{123},H_{456}\}_D=-\frac16$, and the twenty components form ten canonical pairs. The same prescription applied to the fermion term $\frac\ii2\Theta^T\dot\Theta$ gives $\{\Theta_{a\lambda},\Theta_{b\sigma}\}_D=-\ii\delta_{ab}\delta_{\lambda\sigma}$, while the scalars keep their standard brackets with $P=D_tX$. This completes \eqref{dirac}.

\paragraph{The Gauss' law.}
The gauge field appears only through the covariant derivatives, and it acts on all the internal indices. Varying it in the three kinetic terms gives
\begin{eqn}\label{gaussvariation}
 \delta_A L_X&=\delta A_{ab}X_{[a}^IP_{b]}^I,\qquad
 \delta_A L_\Theta=-\frac\ii2\delta A_{ab}\Theta_a^T\Theta_b,\\
 \delta_A L_H&=-\frac32\delta A_{ab}K_{ab} ,
\end{eqn}
where in the last term we used the total antisymmetry of $H$. The sum is $\delta A_{ab}\Ga_{ab}$, which gives the Gauss' law \eqref{gauss}, with the scalar orbital term, the fermion spin and the moment map $K$ of the tensor. As a check of the normalization of $K$, the Dirac bracket of $-\Lambda_{ab}\Ga_{ab}$ with $H$ gives
\begin{eqn}
 \delta H_{abc}=\Lambda_{ad}H_{dbc}+\Lambda_{bd}H_{adc}+\Lambda_{cd}H_{abd},
\end{eqn}
i.e.\ the rotation of a three--form, as expected for the generator of the gauge transformations.

\subsection{Computations on the supersymmetry closure}\label{app:susyclosure}

In this Appendix we derive the results summarized in Section~\ref{sec:mass}, i.e.\ the masses \eqref{massrelations}, the fermion mass and the time dependence of the supersymmetry parameter. The strategy is the Noether procedure used for the plane--wave matrix model \cite{Dasgupta:2002hx,Kim:2006wg}. We start from the massless action \eqref{Lzero}, which is invariant under the $\mu=0$ part of \eqref{susy} \cite{Artime:2026fcr,Artime:2026mph}, we add the most general terms suggested by the plane--wave background, and we require that supersymmetry is preserved. The difference with respect to BMN theory is that, besides the masses, we have two polarization couplings, one for the tensor interaction and one for the five--bracket interaction.

Concretely, we add to the Lagrangian the terms
\begin{eqn}\label{susy:ansatz}
 \Delta L&=-\frac12M_I^2(X_a^I)^2+L_3+L_6-\frac{\ii m_F}{2}\Theta_a^TG\Theta_a,\\
 L_3&=\frac\alpha2\epsilon^{ijk}H_{abc}Y_a^iY_b^jY_c^k,
 \qquad L_6=\beta\det Z .
\end{eqn}
The first term contains the scalar masses, which for the moment we keep independent for each direction. The terms $L_3$ and $L_6$ are the couplings to the three--form flux and to its dual six--form, already introduced in \eqref{Vsquares}, and the last term is a fermion mass. Correspondingly, we modify the fermion transformation by a term linear in the scalars, as in \eqref{massansatz},
\begin{eqn}\label{susy:shift}
 \delta_c\Theta_a=c_IX_a^I\gamma_IG\,\epsilon,
 \qquad c_i=-\alpha,\quad c_m=\beta ,
\end{eqn}
while all the other transformations keep their massless form. The values of $c_I$ are fixed by requiring that $\delta\Theta$ contains the shifted brackets $B-\alpha E_2$ and $C-\beta E_5$ of \eqref{susy}. Indeed, using the identities
\begin{eqn}\label{susy:duality}
 \frac12\epsilon^{ijk}\gamma_{ij}=\gamma_kG,\qquad
 \frac1{5!}\epsilon^{nm_1\ldots m_5}\gamma_{m_1\ldots m_5}=-\gamma_nG ,
\end{eqn}
the shifts $-\alpha E_2$ and $-\beta E_5$ in $\delta\Theta$ become exactly \eqref{susy:shift}. Finally, as we will see, we must also allow the supersymmetry parameter $\epsilon$ to depend on time.

\smallskip

The unknowns are therefore the scalar masses $M_I$, the fermion mass $m_F$, the time dependence of $\epsilon$, and the relation between $\alpha$ and $\beta$. They are fixed in three steps, by collecting the terms of the variation of the action with the same field content:
\begin{enumerate}
\item the terms with one scalar velocity and one fermion $\dot{X} \Theta$ fix the time dependence of $\epsilon$ and the fermion mass;
\item the terms with one scalar and one fermion $X\Theta$ fix the scalar masses $M_I$;
\item the terms with the tensor and two scalars $HX^2\Theta$, and with five scalars $FX^5\Theta$, fix the relation $\alpha=2\beta$.
\end{enumerate}

\subsubsection*{Time dependence of $\epsilon$ and fermion mass}

In the massless theory the parameter $\epsilon$ is constant to enforce supersymmetry. Indeed, if we allow it to depend on time, the massless action is no longer invariant, and its variation is proportional to $\dot\epsilon$,
\begin{eqn}\label{susy:kterm}
 \delta_0L_0\simeq-\ii\dot\epsilon^TQ_0,
\end{eqn}
where $Q_0$ is the charge \eqref{Q} at $\mu=0$, and $\simeq$ denotes equality up to a total time derivative. However, in the massive $\mathcal{A}_6$ model the fermion mass and the shift $\delta_c\Theta$ produce new terms proportional to the scalar velocities, and \eqref{susy:kterm} is precisely what is needed to cancel them.

To treat the two target blocks at once, we note that $\gamma_I$ commutes with $G=\gamma_{123}$ if $I$ is in the three--plane, and anticommutes with it if $I$ is in the six--plane. We write this as
\begin{eqn}\label{susy:signs}
 G\gamma_I=s_I\gamma_IG,\qquad s_i=+1,\quad s_m=-1,
\end{eqn}
so that $(\delta_c\Theta_a)^T=-c_Is_IX_a^I\epsilon^T\gamma_IG$. We now collect the terms with one velocity and one fermion. The first comes from the fermion kinetic term, which after an integration by parts gives
\begin{eqn}\label{susy:kinvariation}
 \delta_cL_{\mathrm{kin},F}
 &\simeq-\ii c_Is_IX_a^I\epsilon^T\gamma_IGD_t\Theta_a\\
 &\simeq\ii c_Is_IP_a^I\epsilon^T\gamma_IG\Theta_a
       +\ii c_Is_IX_a^I\dot\epsilon^T\gamma_IG\Theta_a .
\end{eqn}
The second comes from the fermion mass term,
\begin{eqn}\label{susy:massvelocity}
 \delta_0L_{F,\mathrm{mass}}\big|_P
 =-\ii m_FP_a^I\epsilon^T\gamma_IG\Theta_a ,
\end{eqn}
and the third is the momentum part of \eqref{susy:kterm}. Requiring that the coefficient of each $P_a^I\Theta_a$ vanishes gives
\begin{eqn}\label{susy:parameterrow}
 \dot\epsilon^T\gamma_I=(c_Is_I-m_F)\epsilon^T\gamma_IG .
\end{eqn}
Multiplying by $\gamma_I$ and using $\gamma_IG\gamma_I=s_IG$, this becomes a differential equation for $\epsilon$,
\begin{eqn}\label{susy:parameter}
 \dot\epsilon=(m_Fs_I-c_I)\,G\epsilon\equiv kG\epsilon .
\end{eqn}
The important point is that $\epsilon$ is the same spinor for all directions, so the coefficient $k$ must be the same in the three--plane and in the six--plane. This gives $k=m_F+\alpha=-m_F-\beta$, and therefore
\begin{eqn}\label{susy:km}
 k=\frac{\alpha-\beta}{2},\qquad m_F=-\frac{\alpha+\beta}{2} .
\end{eqn}
The solution is $\epsilon(t)=e^{ktG}\epsilon_0$. Since $G^2=-1$, this is a rotation of the sixteen supersymmetry parameters with frequency $k$, which vanishes only if $\alpha=\beta$.

\medskip

We now collect the terms with one scalar and one fermion. Using \eqref{susy:parameter}, the second term of \eqref{susy:kinvariation} becomes $\ii kc_IX_a^I\epsilon^T\gamma_I\Theta_a$. Adding the variation of the scalar mass term and the variation of the fermion mass under $\delta_c\Theta$, we find
\begin{eqn}\label{susy:scalarmasses}
 \delta L\big|_{X\Theta}
 &=\ii\bigl(M_I^2+kc_I-m_Fc_Is_I\bigr)X_a^I\epsilon^T\gamma_I\Theta_a\\
 &=\ii(M_I^2-c_I^2)X_a^I\epsilon^T\gamma_I\Theta_a ,
\end{eqn}
where in the second line we used $m_Fs_I=c_I+k$, which follows from \eqref{susy:parameter}. Hence supersymmetry requires $M_I^2=c_I^2$, i.e.\ $M_i=\alpha$ and $M_m=\beta$. These are exactly the masses produced by completing the squares in \eqref{Vsquares}, which is a first consistency check of the construction, but trivially expected.

\subsubsection*{The tensor interactions}

We now consider the terms with one tensor, two scalars and one fermion, i.e.\ of the form $HXX\Theta$. In the massless theory these terms already cancel, so we only need to collect the new contributions produced by the deformation. There are four of them.

The first one comes from the fermion mass, varied with the two--bracket part of $\delta_0\Theta$,
\begin{eqn}\label{susy:Bmass}
 \delta_BL_{F,\mathrm{mass}}
 =\frac{\ii m_F}{2}B_a^{IJ}\epsilon^T\gamma_{IJ}G\Theta_a .
\end{eqn}
The second one comes from the tensor Yukawa coupling, varied with the shift $\delta_c\Theta$. Since $W_H$ is antisymmetric in the combined internal and spinor indices, the variations of the two fermions give the same contribution, and using \eqref{susy:signs} we find
\begin{eqn}\label{susy:WH}
 \delta_cL_{Y,H}
 &=\ii H_{abc}X_c^I(\delta_c\Theta_a)^T\gamma_I\Theta_b\\
 &=-\ii H_{abc}c_JX_a^JX_c^I\epsilon^TG\gamma_J\gamma_I\Theta_b\\
 &=\frac\ii2H_{abc}X_b^IX_c^J\epsilon^T(c_I+c_J)G\gamma_{IJ}\Theta_a .
\end{eqn}
The third contribution comes from the time dependence of $\epsilon$. Indeed, the two--bracket part of \eqref{susy:kterm} gives $-\frac\ii2kB_a^{IJ}\epsilon^TG\gamma_{IJ}\Theta_a$. Finally, the fourth one comes from the polarization term $L_3$, whose three scalars give three equal terms,
\begin{eqn}\label{susy:cubicY}
 \delta_YL_3
 =-\frac{3\ii\alpha}{2}\epsilon^{nIJ}H_{abc}X_b^IX_c^J
       \epsilon^T\gamma_n\Theta_a ,
\end{eqn}
which are nonzero only in the three--plane.

\smallskip

Collecting the four contributions, the variation takes the form
\begin{eqn}\label{susy:MIJ}
 \delta L\big|_{HX^2\Theta}
 &=\frac\ii2H_{abc}X_b^IX_c^J\epsilon^T\mathcal A_{IJ}\Theta_a,\\
 \mathcal A_{IJ}
 &=m_F\gamma_{IJ}G+(c_I+c_J-k)G\gamma_{IJ}
                  -3\alpha\epsilon^{nIJ}\gamma_n ,
\end{eqn}
and supersymmetry requires $\mathcal A_{IJ}=0$. Since $G$ commutes with $\gamma_{IJ}$ when both indices are in the same plane, and anticommutes with it when they are in different planes, we have to distinguish three cases. Using \eqref{susy:km} we find
\begin{eqn}\label{susy:pairclasses}
 \mathcal A_{im}&=(-m_F-\alpha+\beta-k)G\gamma_{im}
                  =(2\beta-\alpha)G\gamma_{im},\\
 \mathcal A_{mn}&=(m_F+2\beta-k)G\gamma_{mn}
                  =(2\beta-\alpha)G\gamma_{mn},\\
 \mathcal A_{ij}&=-3\alpha G\gamma_{ij}-3\alpha\epsilon^{nij}\gamma_n=0 ,
\end{eqn}
where in the last line we used $G\gamma_{ij}=-\epsilon^{nij}\gamma_n$. The three--plane terms therefore cancel identically, thanks to the polarization term $L_3$, while the other two cases vanish only if
\begin{eqn}
 \alpha=2\beta .
\end{eqn}
This is the relation between the deformations of the two planes anticipated in Section~\ref{sec:mass}. Together with \eqref{susy:km}, it fixes all the parameters in terms of a single scale, giving \eqref{massrelations}.

\subsubsection*{The five--bracket interaction}

Finally, we consider the terms with five scalars and one fermion, which contain the five--bracket $C_a^{IJKLM}$. The logic is the same as for the tensor sector, but the bookkeeping is simpler if we classify the five--bracket by the number $r$ of its directions lying in the three--plane. Since the three--plane has only three directions, $r$ can take the values $0,1,2,3$, and the remaining $5-r$ directions lie in the six--plane.

For each $r$, the contributions of the fermion mass, of the time dependence of $\epsilon$ and of the shift $\delta_c\Theta$ in the five--bracket Yukawa coupling combine into a single term proportional to
\begin{eqn}
 \frac{\ii}{5!}\,A_r\,C_a^{IJKLM}\epsilon^TG\gamma_{IJKLM}\Theta_a .
\end{eqn}
The coefficient $A_r$ depends on $r$ in two ways. Moving $G$ through the five gamma matrices gives a sign $(-1)^{r+1}$, and the shift $\delta_c\Theta$ contributes the sum of the coefficients $c_I$ over the five directions, which is $-r\alpha+(5-r)\beta$. Using \eqref{susy:km}, we obtain
\begin{eqn}\label{susy:Ar}
 A_r=-m_F(-1)^{r+1}+k+r\alpha-(5-r)\beta,\qquad\\
 (A_0,A_1,A_2,A_3)=(-6\beta,\,2\alpha-4\beta,\,2\alpha-4\beta,\,4\alpha-2\beta).
\end{eqn}
These coefficients do not vanish by themselves, and they must be cancelled by the two polarization terms. Each of them acts only on one value of $r$:
\begin{itemize}
\item for $r=0$, i.e.\ all directions in the six--plane, the variation of $L_6=\beta\det Z$ contributes $+6\beta$;
\item for $r=3$, i.e.\ the full three--plane together with two directions of the six--plane, the variation of $H$ in $L_3$ contributes $-3\alpha$.
\end{itemize}
The total coefficients are therefore
\begin{eqn}\label{susy:fiveclasses}
 r=0:\ 0,\qquad r=1:\ 2(\alpha-2\beta),\qquad r=2:\ 2(\alpha-2\beta),\qquad r=3:\ \alpha-2\beta ,
\end{eqn}
and they all vanish for $\alpha=2\beta$, the same relation found in the tensor sector. This is a non--trivial consistency check of the construction, since the two sectors involve different couplings.

Finally, all the new terms in the variation contain a single fermion, and the cubic fermion terms are the same as in the massless theory, since the shift $\delta_c\Theta$ is linear in the scalars. Hence the conditions found above are all the conditions imposed by supersymmetry, and with $\mu=3\alpha$ they give \eqref{massrelations}, the action \eqref{Lmass} and the transformations \eqref{susy}.

\subsection{The canonical supersymmetry algebra}\label{app:chargealgebra}

In this part we derive the algebra \eqref{algebra} of the supersymmetry charges. Let us first explain what we would expect from this computation. The charge $Q$ in \eqref{Q} generates the supersymmetry transformations through the Dirac brackets \eqref{dirac}. Computing $\{Q_\lambda,Q_\sigma\}_D$ tells us which bosonic symmetries are generated by two supersymmetry transformations. In flat space one finds only the Hamiltonian and a gauge transformation. In the plane--wave we expect also the rotations of the two planes, as in BMN theory \cite{Dasgupta:2002hx}.

\medskip

We split the charge into its massless part and the part produced by the deformation,
\begin{eqn}\label{app:Qsplit}
 Q=Q_0+Q_\mu,\qquad
 Q_0=\sum_aN_a\Theta_a,\qquad Q_\mu=\sum_aD_a\Theta_a ,
\end{eqn}
where $N_a$ and $D_a$ are $16\times16$ matrices which depend only on the bosonic fields,
\begin{eqn}\label{app:ND}
 N_a&=P_a^I\gamma_I-\frac12B_a^{IJ}\gamma_{IJ}
     +\frac1{5!}C_a^{IJKLM}\gamma_{IJKLM},\\
 D_a&=\alpha Y_a^i\gamma_iG+\beta Z_a^m\gamma_mG .
\end{eqn}
The bracket then splits into three parts,
\begin{eqn}
 \{Q,Q\}_D=\{Q_0,Q_0\}_D+\bigl(\{Q_0,Q_\mu\}_D+\{Q_\mu,Q_0\}_D\bigr)+\{Q_\mu,Q_\mu\}_D ,
\end{eqn}
i.e.\ the massless algebra, the cross terms and the square of the deformation. Moreover, each bracket receives two kinds of contributions. Since $Q$ is linear in $\Theta$, the bracket of the two fermions, $\{\Theta_{a\lambda},\Theta_{b\sigma}\}_D=-\ii\delta_{ab}\delta_{\lambda\sigma}$, gives a term which contains only bosons. The brackets of the bosonic coefficients $N_a$ and $D_a$, instead, leave the two fermions untouched, and give a fermion bilinear. We now compute these contributions one by one.

\paragraph{The massless part.}
Contracting the fermions in $\{Q_0,Q_0\}_D$ gives the matrix $-\ii\sum_aN_aN_a^T$. Squaring \eqref{app:ND}, the part proportional to the identity contains $P^2$ and the squares of the two brackets, i.e.\ twice the bosonic massless Hamiltonian, while the remaining terms reconstruct the bosonic Gauss' law multiplied by $W_+$,
\begin{eqn}\label{app:QQbosonic}
 \sum_aN_aN_a^T=2\Ham_{B,0}\mathbf1-2\Ga_{B,ab}(W_+)_{ab} .
\end{eqn}
The brackets of the coefficients $N_a$, computed with \eqref{dirac}, give instead the fermionic terms,
\begin{eqn}\label{Ridentity}
 \sum_{a,I}\bigl(\partial_{X_a^I}Q_0\,\partial_{P_a^I}Q_0
 -\partial_{P_a^I}Q_0\,\partial_{X_a^I}Q_0\bigr)
 -\frac16J_{tu}\,\partial_{H_t}Q_0\,\partial_{H_u}Q_0
 =-(\Theta^TW_-\Theta)\,\mathbf1+\Theta_a^T\Theta_b(W_+)_{ab} .
\end{eqn}
The first term is the Yukawa part of the Hamiltonian, and the second is the fermion part of the Gauss' law. Notice that the tensor contributes through its Dirac bracket, i.e.\ the last term on the left--hand side. Adding \eqref{app:QQbosonic} and \eqref{Ridentity}, we recover the massless algebra of \cite{Artime:2026fcr,Artime:2026mph}, $\{Q_0,Q_0\}_D=-2\ii\Ham_0+2\ii\Ga W_+$.\footnote{To rearrange the spinor indices in \eqref{Ridentity} one expands each $16\times16$ matrix on the basis of antisymmetrized gamma matrices of rank zero to four, which contains $1+9+36+84+126=256$ elements,
\begin{eqn}\label{cliffcomplete}
 \delta_{\lambda\nu}\delta_{\sigma\tau}
 =\frac1{16}\sum_{k=0}^{4}\frac{(-1)^{k(k-1)/2}}{k!}
 (\gamma_{I_1\ldots I_k})_{\lambda\sigma}
 (\gamma_{I_1\ldots I_k})_{\tau\nu} ,
\end{eqn}
and reduces products with $\gamma_I\gamma_{J_1\ldots J_k}=\gamma_{IJ_1\ldots J_k}+\sum_{r}(-1)^{r-1}\delta_{IJ_r}\gamma_{J_1\ldots\widehat J_r\ldots J_k}$.}

\paragraph{The pure deformation part.}
Since $D_a$ contains no momenta, the bosonic brackets in $\{Q_\mu,Q_\mu\}_D$ vanish, and only the fermion contraction contributes. Using $G^2=-1$ we find $\sum_aD_aD_a^T=(\alpha^2Y^2+\beta^2Z^2)\,\mathbf1$, which is twice the mass terms of the scalars.

\paragraph{The cross terms.}
These are the most interesting ones, since they produce the new terms of the algebra. Contracting the fermions gives the matrix $\sum_a(N_aD_a^T+D_aN_a^T)$, which contains two types of terms. The products of $D_a$ with the two brackets in $N_a$ give, after using $\alpha=2\beta$,
\begin{eqn}\label{app:QQpolarization}
 -\bigl(\alpha\epsilon^{ijk}H_{abc}Y_a^iY_b^jY_c^k+2\beta\det Z\bigr)\,\mathbf1 ,
\end{eqn}
i.e.\ twice the polarization terms of \eqref{Vsquares}. Together with the square of the deformation, this completes the bosonic Hamiltonian of the massive model. The products of $D_a$ with the momentum term of $N_a$ give instead the rotations. Indeed, for $I\neq J$ one finds
\begin{eqn}\label{app:QQrotationidentity}
 \gamma_I(D\text{--coefficient of }X^J)^T+(D\text{--coefficient of }X^J)\gamma_I=
 \begin{cases}
 -2\alpha\,\gamma_{ij}G, & I,J \text{ in the three--plane},\\
 +2\beta\,\gamma_{mn}G, & I,J \text{ in the six--plane},\\
 0, & \text{otherwise},
 \end{cases}
\end{eqn}
and multiplying by $P^I_aX^J_a$ gives the orbital angular momenta $X^IP^J-X^JP^I$ of the two planes, with the coefficients of \eqref{algebra}. Finally, the bosonic brackets between $D_a$ and the momenta in $N_a$ leave a fermion bilinear,
\begin{eqn}\label{app:QQmassspin}
 -\frac\mu4(\Theta^TG\Theta)\,\mathbf1
 -\frac\alpha4(\gamma_{ij}G)\,\Theta^T\gamma_{ij}\Theta
 +\frac\beta4(\gamma_{mn}G)\,\Theta^T\gamma_{mn}\Theta .
\end{eqn}
The first term is the fermion mass in the Hamiltonian, and the other two are the spin parts of the rotations \eqref{targetJ}.

\medskip

Collecting all the contributions, the Hamiltonian appears complete, with its masses, polarization terms and fermion mass, together with the Gauss' law and the full angular momenta of the two planes. This gives exactly \eqref{algebra}.

\section{Computations on the plane--wave M5--brane theory}\label{app:superM5}
We derive the projected superfields used in Section~\ref{sub:LCM5}. The calculation follows the normal-coordinate construction of \cite{Claus:1998fh}, applied to the plane--wave as in the membrane analysis of \cite{Dasgupta:2002hx}. Its advantage is that the fermionic expansion follows from the supergravity torsion and curvature constraints, so the same background superfields can be used in both the M2 and M5 actions.

\subsubsection*{Clifford representation and the flux connection}
To impose $\G^+\vartheta=0$ explicitly, choose the real thirty-two-dimensional matrices
\begin{eqn}\label{app:cliffordblocks}
 \G^+=\sqrt2\begin{pmatrix}0&0\\1&0\end{pmatrix},\quad
 \G^-=\sqrt2\begin{pmatrix}0&-1\\0&0\end{pmatrix},\quad
 \G^I=\begin{pmatrix}\gamma_I&0\\0&-\gamma_I\end{pmatrix},\quad
 \mathsf C=\begin{pmatrix}0&-1\\1&0\end{pmatrix}.
\end{eqn}

where each block is sixteen-dimensional and $\mathsf C$ is the charge-conjugation matrix. The gauge condition sets the upper block of $\vartheta$ to zero, giving
\begin{eqn}\label{app:projectedspinor}
 \vartheta=2^{-1/4}\begin{pmatrix}0\\\psi\end{pmatrix},\qquad
 \vartheta^T\mathsf C=2^{-1/4}(\psi^T,0),\qquad
 \G^+\vartheta=0,\qquad \vartheta^T\mathsf C\G^+=0 .
\end{eqn}
The normalization ensures that $\vartheta^T\mathsf C\G^-d\vartheta=-\psi^Td\psi$, which will give the fermionic part of $E^-$.\footnote{We use a left exterior derivative and the exchange sign $(-1)^{pq+fg}$ for forms of degree $p,q$ and intrinsic Grassmann parity $f,g$. In particular, the odd one-forms $d\vartheta$ commute under this combined grading.}
The superspace constraints of \cite{Claus:1998fh,Bergshoeff:1998vx}, translated to these exterior and Clifford conventions, are
\begin{eqn}\label{superconstraints}
 T^a&=\ii E^T\mathsf C\G^aE,\\
 F_4&=F_{4,B}+\frac\ii2 E^T\mathsf C\G_{ab}E\wedge E^a\wedge E^b,\\
 F_7&=F_{7,B}-\frac\ii{5!}E^T\mathsf C\G_{a_1\ldots a_5}E
                         \wedge E^{a_1}\wedge\cdots\wedge E^{a_5},\\
 F_4&=dC_3,\qquad F_7=dC_6-\tfrac12C_3\wedge F_4,\qquad
 \mathscr H=dA_2-C_3 .
\end{eqn}
Here $E$ without a vector index denotes the spinor vielbein. 
The spinor derivative in the normal-coordinate construction contains both the metric connection and the four-form flux \cite{Claus:1998fh,Dasgupta:2002hx},
\begin{eqn}\label{fluxconnection}
 \mathcal D\vartheta&=(d+\tfrac14\omega_{ab}\G^{ab}+e^a\Omega_a)\vartheta,\\
 \Omega_a=-\frac1{288}(\G_a\G^{bcde}F_{bcde}-&12F_{abcd}\G^{bcd}) = -\frac{1}{288}(\G_a{}^{bcde}F_{bcde}-8F_{abcd}\G^{bcd}).
\end{eqn}
To compute such term, note that for the null frame $e^+=dt$, $e^-=dx^-+Udt/2$, and $e^I=dX^I$, the spin connection has an electric component proportional to $\partial_IU\,\G^{+I} \propto \partial_IU \G^I \G^+$ which annihilates the projected spinor in this gauge choice.
To evaluate the remaining flux connection, set $F_{+123}=-\mu$. The contractions present in \eqref{fluxconnection} are
\begin{eqn}\label{app:fluxcontractions}
 \G^{bcde}F_{bcde}=-24\mu\G^{+123},\qquad
 F_{+bcd}\G^{bcd}=-6\mu\G^{123} .
\end{eqn}
Using $\G_+=-\G^-$ in \eqref{fluxconnection} gives
\begin{eqn}\label{app:Omegapluscalculation}
 \Omega_+=\frac\mu{12}\G_+\G^{+123}-\frac\mu4\G^{123}
 =\begin{pmatrix}-\mu G/12&0\\0&\mu G/4\end{pmatrix} .
\end{eqn}
The transverse connection terms contain $\G^+$ on the projected block, and the minus component vanishes. Thus
\begin{eqn}\label{app:fluxblocks}
 \Omega_+|_{\mathrm{upper}}=-\mu G/12,\qquad
 \Omega_+|_{\mathrm{lower}}=\mu G/4,\qquad
 \Omega_I\vartheta=\Omega_-\vartheta=0 \,,
\end{eqn}
and the surviving derivative is just $\mathcal D\psi=d\psi+\mu G\psi\,dt/4$. 

\paragraph{Comments on the Killing Spinor Equations.} On the other side, the KSE $\delta_K \Psi_M=0$ contain 
\begin{eqn}
    \left( \partial_M + \frac{1}{4} \omega_{Mab}\G^{ab} + e_M^a \Omega_a \right) \epsilon_K =0\,.
\end{eqn}
On the transverse origin $X^I = 0$, the plane--wave satisfies $U(0) = \partial_I U(0) = 0$. In this case the time component of the KSE reads as 
\begin{eqn}
    (\partial_t + \Omega_+)\epsilon_K(t)=0\,,
\end{eqn}
where in general $\epsilon_K(t) = (\epsilon(t), \eta(t))^T$. Hence, the upper block of the Killing-spinor equation instead reads $\partial_t\epsilon-\mu G\epsilon/12=0$ at the transverse origin, as obtained in \eqref{susy}.\footnote{
Away from the origin, the full Killing spinor also has its usual transverse-coordinate dependence, while this upper--block parameter still determines the dynamical transformations in light-cone gauge.} 
Instead the lower-block equation gives
\begin{eqn}
    \eta(t)=e^{-\mu tG/4}\eta_0\,,
\end{eqn}
which generates a spatially constant shift of the worldvolume
fermion. In the spherical reduction, this transformation acts
on the constant harmonic $\Theta_0$ and leaves the first-harmonic
coefficients $\Theta_a$ unchanged. It therefore reproduces the
sixteen kinematical supersymmetries of the center-of-mass sector
introduced in Section~\ref{sec:mass}, extending the constant-mode construction
of \cite{Artime:2026fcr,Artime:2026mph} to the plane--wave.

Instead the potentials follow by integrating their curvatures along the same fermionic path, as in \cite{Claus:1998fh}. The contraction of a curvature with the fermionic radial direction removes one spinor differential. In the ordering adopted here, the bilinear primitives are
\begin{eqn}\label{app:primitives}
 (C_3-C_{3,B})_{\vartheta^2}
 &=\frac{\ii}{2}\vartheta^T\mathsf C\G_{ab}\mathcal D\vartheta
                       \wedge e^a\wedge e^b,\\
 (C_6-C_{6,B})_{\vartheta^2}
 &=-\frac{\ii}{5!}\vartheta^T\mathsf C\G_{a_1\ldots a_5}\mathcal D\vartheta
                       \wedge e^{a_1}\wedge\cdots\wedge e^{a_5}.
\end{eqn}

The required nonzero spinor blocks are
\begin{eqn}\label{app:projectedbilinears}
 \vartheta^T\mathsf C\G_{+I}d\vartheta=-\psi^T\gamma_Id\psi,
 \qquad
 \vartheta^T\mathsf C\G_{+IJKL}d\vartheta=+\psi^T\gamma_{IJKL}d\psi .
\end{eqn}
These identities follow by multiplying \eqref{app:cliffordblocks}. The lower plus index is $\G_+=-\G^-$, so it connects the lower physical spinor to the nonzero upper row in \eqref{app:projectedspinor}, while purely transverse components and components with a lower minus index vanish.\footnote{
In the three-form primitive, the two possible positions of the plus index cancel the factor $1/2$, leaving $-\ii\,dt\wedge dX^I\wedge\psi^T\gamma_Id\psi$. In the six-form primitive, the five positions of the plus index turn $1/5!$ into $1/4!$. Moving the fermionic one-form past the five bosonic one-forms gives the additional minus sign, leaving the positive bilinear in \eqref{exactpotentials}.}

\subsection{Canonical reduction and the fermionic terms}\label{app:canonicalM5}

In this section we derive the light--cone action \eqref{fullM5}. We start from the Hamiltonian formulation of the M5--brane in a supergravity background \cite{Bergshoeff:1998vx}, and we impose the light--cone gauge as in the flat--space analysis of \cite{Bandos:2008fr}. The derivation has three steps. We first compute how the background potentials enter the action and then we fix the light--cone gauge and solve the constraints, which gives the bosonic part of \eqref{fullM5}. Finally we add the fermions, whose couplings follow from the superfields \eqref{exactvielbein} and \eqref{exactpotentials}.

\paragraph{The coupling to the background potentials.}
In the Hamiltonian formulation, the background potentials enter by shifting the canonical momenta of the embedding coordinates, as the vector potential shifts the momentum of a charged particle. In our conventions the shift is \cite{Bergshoeff:1998vx}
\begin{eqn}\label{M5shift}
 \widehat C_M=\star_\Sigma\left[
 \iota_M C_6-\frac12(\iota_M C_3)\wedge(C_3+2\mathscr H)
 \right]_\Sigma ,
\end{eqn}
where $\star_\Sigma$ extracts the density of a five--form on $\Sigma$. The plane--wave potentials \eqref{exactpotentials} have the form $C_3=dt\wedge c_2$ and $C_6=dt\wedge c_5$, i.e.\ they always contain $dt$. Since $x^+=t$, only the contraction with $\partial_+$ removes this $dt$ and leaves a form with a non--vanishing pullback to $\Sigma$. Using $\mathscr H|_\Sigma=h$, we therefore find
\begin{eqn}\label{app:shiftcomponents}
 \widehat C_I=\widehat C_\alpha=\widehat C_-=0,\qquad
 \widehat C_+=\star_\Sigma(c_5-h\wedge c_2).
\end{eqn}
Hence the background only shifts the light--cone energy (as expected), by the coupling to the six--form and by a coupling of the tensor to the three--form. The factor $2\mathscr H$ in \eqref{M5shift} is the sum of two contributions, one from the momentum of the tensor and one from the Wess--Zumino term, and it gives the coupling $-T_5\int_\Sigma h\wedge c_2$ of \eqref{fullM5}.

\paragraph{The bosonic light--cone action.}
In the light--cone gauge $E_i^+=0$ and $C_3|_\Sigma=0$, so the spatial metric and the spatial field strength are simply
\begin{eqn}
 g_{ij}=\partial_iX^I\partial_jX^I,\qquad h=d_\Sigma A_2 .
\end{eqn}
After imposing the constraint of the spatial diffeomorphisms, the energy of the fivebrane contains two terms, the volume of $\Sigma$ and a term quadratic in the flux \cite{Bergshoeff:1998vx,Bandos:2008fr}. Both can be written with the brackets of Section~\ref{sub:LCM5},
\begin{eqn}\label{app:geometricidentities}
 \det g&=\frac{w^2}{5!}\sum_{I_1,\ldots,I_5}
              \N(X^{I_1},\ldots,X^{I_5})^2,\\
 g_{ik}g_{jl}b^{ij}b^{kl}
       &=\sum_{I,J}\Bh(X^I,X^J)^2 .
\end{eqn}
This is why the volume energy and the tensor energy become the squares of the two brackets in \eqref{fullM5}. Finally, at fixed light--cone momentum $P_-=-pw$, the equation for the transverse momenta gives $\Pi_I=pD_tX^I$, and substituting it back produces the scalar kinetic term, while the tensor keeps its first--order kinetic term $\tfrac{T_5}{2}\int_\Sigma D_tA_2\wedge h$.

\paragraph{The fermions.}
The fermions enter in two ways. The first is through the longitudinal vielbein, whose fermionic part in \eqref{exactvielbein}. At fixed $P_-=-pw$, the coefficient of $E^-$ is $p$, so these terms give the kinetic term and the mass of the fermions,
\begin{eqn}
 \ii p\int_\Sigma w\,\psi^T D_t\psi
 +\frac{\ii\mu p}{4}\int_\Sigma w\,\psi^T G\psi .
\end{eqn}
The second is through the fermionic parts of the potentials \eqref{exactpotentials}, which we denote by $c_{2,F}$ and $c_{5,F}$, so that $c_r=c_{r,B}+c_{r,F}$. Inserting them in \eqref{app:shiftcomponents}, we find the two Yukawa couplings,
\begin{eqn}\label{app:yukawaorigins}
 -T_5\int_\Sigma h\wedge c_{2,F}
 &=\ii T_5\int_\Sigma w\,\psi^T\gamma_I\Bh(X^I,\psi),\\
 T_5\int_\Sigma c_{5,F}
 &=\frac{\ii T_5}{4!}\int_\Sigma w\,\psi^T\gamma_{IJKL}
        \N(X^I,X^J,X^K,X^L,\psi).
\end{eqn}
In the first line, the contraction with $h$ produces the bivector $b^{ij}$, and hence the bracket $\Bh$ between a scalar and the fermion. In the second line, the four differentials $dX^I$ and the differential $d\psi$ combine into the Nambu bracket of four scalars and the fermion. Both Yukawa couplings are already present in flat space \cite{Artime:2026mph}. The only new fermionic term produced by the plane--wave is therefore the mass, which comes from $E^-$.

\subsection{Sphere reduction of the action}\label{app:sphereM5}

In this Appendix we show that the spherical ansatz \eqref{M5dictionary} reduces the action \eqref{fullM5} to \eqref{actionidentity}. The massless part of the computation was done in \cite{Artime:2026mph}, and here we recall it briefly, in order to fix the normalization of the new plane--wave terms. The strategy is simple. Every term of \eqref{fullM5} is an integral over the sphere of products of the fields, so we only need two ingredients: the integral of two linear functions, and the action of the brackets on linear functions.

\paragraph{Two basic identities.}
In the frame of \eqref{dictionary} we have
\begin{eqn}\label{app:spheremoments}
 \int_{\mathbf{S}^5 }w\,u_au_b=\frac{\Omega_5}{6}\delta_{ab},\qquad
 \N(u_b,u_c,u_d,u_e,u_f)=-\epsilon_{abcdef}u_a .
\end{eqn}
The first identity follows from rotational invariance. The integral must be proportional to $\delta_{ab}$, and since $u_au_a=1$ its trace is $\Omega_5$, which fixes the coefficient. The second identity is the Nambu bracket of the coordinates of the sphere \cite{Artime:2026mph}, with the orientation used throughout this paper.

\paragraph{The brackets on the ansatz.}
On the ansatz, the field strength is $h=-6\,(*_6\mathsf H)|_{\mathbf{S}^5 }$, i.e.\ the restriction to the sphere of the constant three--form of $\mathbb{R}^6$ dual to $\mathsf H$, as in \cite{Artime:2026mph}. Inserting it in \eqref{tensorbivector} and using \eqref{app:spheremoments}, we find
\begin{eqn}\label{operatorreduction}
 \Bh(u_b,u_c)&=-6H_{abc}u_a,\\
 \gamma_I\Bh(X^I,\psi)&=\frac6{\sqrt2}\,u_a(W_H)_{ab}\Theta_b,\\
 \frac{1}{4!}\gamma_{IJKL}\N(X^I,X^J,X^K,X^L,\psi)
 &=-\frac6{\sqrt2}\,u_a(W_F)_{ab}\Theta_b .
\end{eqn}
The second line follows from the first, inserting $X^I=X_b^Iu_b$ and $\psi=\Theta_cu_c/\sqrt2$ and exchanging two indices of $H$. The third line follows in the same way from the Nambu bracket in \eqref{app:spheremoments}, using $\epsilon_{abcdef}=6F_{abcdef}$. In both cases we obtain the Yukawa operators $W_H$ and $W_F$ of \eqref{BCW}.

\paragraph{The massless terms.}
We can now evaluate the action term by term. Since all the fields are linear in the $u_a$, every quadratic term reduces with the first identity of \eqref{app:spheremoments}. For the kinetic terms this gives
\begin{eqn}\label{app:kineticmatch}
 \frac p2\int w(D_tX)^2&=\frac{\mathcal Z}{2}(D_tX_a)^2,\qquad
 \ii p\int w\,\psi^TD_t\psi=\frac{\ii\mathcal Z}{2}\Theta_a^TD_t\Theta_a,
\end{eqn}
again with $\mathcal Z=p\Omega_5/6$. The interaction terms reduce in the same way, after using \eqref{operatorreduction}. Finally, the kinetic term of the tensor is not modified by the plane--wave, and its reduction is the one of \cite{Artime:2026mph},
\begin{eqn}\label{app:tensorkinetic}
 \frac{T_5}{2}\int_{\mathbf{S}^5 }D_tA_2\wedge h
 =\frac{\mathcal Z}{2}H_{abc}F_{abcdef}D_tH_{def} ,
\end{eqn}
where exact contributions to $D_tA_2$ integrate to zero on the closed sphere, since $d_\Sigma h=0$.

\paragraph{The plane--wave terms.}
The new terms are the masses and the polarization couplings. The masses are quadratic, so again only the first identity of \eqref{app:spheremoments} is needed,
\begin{eqn}\label{app:massmatch}
 -\frac p2\int wU&=-\frac{\mathcal Z}{2}
 \left[\Bigl(\frac\mu3\Bigr)^2(Y_a^i)^2
      +\Bigl(\frac\mu6\Bigr)^2(Z_a^m)^2\right],\qquad
 \frac{\ii\mu p}{4}\int w\,\psi^TG\psi
 =\frac{\ii\mu\mathcal Z}{8}\Theta_a^TG\Theta_a .
\end{eqn}
For the polarization couplings we write the three--plane volume as $\iota_Y\vol_3=\frac12\epsilon^{ijk}Y^i\,dY^j\wedge dY^k$, and since $dY^j\wedge dY^k\wedge h=w\,\Bh(Y^j,Y^k)\,d^5\sigma$, the first line of \eqref{operatorreduction} gives
\begin{eqn}\label{app:electricintegrals}
 \int_{\mathbf{S}^5 }(-h\wedge c_{2,B})
 &=\frac{\mu\Omega_5}{6}\epsilon^{ijk}H_{abc}Y_a^iY_b^jY_c^k,\qquad
 \int_{\mathbf{S}^5 }c_{5,B}=\frac{\mu\Omega_5}{6}\det Z ,
\end{eqn}
where the second integral follows in the same way from the Nambu bracket. Multiplying by $T_5/\mathcal Z$ and using $6T_5/p=1$, we obtain \eqref{electricreduction}, with $\det Z=E_5\cdot C/5!$ as in \eqref{detflux}. Collecting all the terms, the fivebrane action on the spherical ansatz is 
\begin{eqn}
   \left.S_{\mathrm{M5}}\right|_{\mathrm{spherical\ modes}}=
       \mathcal Z\int dt\,L_\mu \,,
\end{eqn}
as in \eqref{actionidentity}.

\bibliographystyle{A6JHEP}
\bibliography{A6_massive}
\end{document}